\documentclass[preprint,aps,amssymb,showpacs,superscriptaddress,nofootinbib]{revtex4}
\usepackage{graphicx}
\newcommand{\del}{\partial}

\def\ee{\end{equation}}
\def\beq{\begin{eqnarray}}
\def\eeq{\end{eqnarray}}
\def\bn{\begin{eqnarray*}}
\def\en{\end{eqnarray*}}
\def\w{\omega}
\def\g{\gamma}
\def\S{\Sigma}

\def\a{\alpha}
\def\b{\beta}
\def\m{\mu}
\def\n{\nu}

\def\th{\theta}

\def\e{\epsilon}

\def\draftversion{Y}                
\def\note[#1]#2{\message{(#1)}\if\draftversion
{\noindent\em[#2]\/}\fi}
\if\draftversion N
\fi

\begin{document}

\title{Topological parameters in gravity }

\author{Romesh K. Kaul}
\email{kaul@imsc.res.in}
\affiliation{The Institute of Mathematical Sciences\\ 
CIT Campus, Chennai-600 113, INDIA.}

\author{Sandipan Sengupta}
\email{sandi@imsc.res.in}
\affiliation{The Institute of Mathematical Sciences\\
CIT Campus, Chennai-600 113, INDIA.}

\begin{abstract}
  We present the Hamiltonian analysis of the theory of gravity based
  on a Lagrangian density containing Hilbert-Palatini term along with
  three topological densities, Nieh-Yan, Pontryagin and Euler.  The
  addition of these topological terms modifies the symplectic
  structure non-trivially. The resulting canonical theory develops a
  dependence on three parameters which are coefficients of these
  terms. In the time gauge, we obtain a real $SU(2)$ gauge theoretic
  description with a set of seven first class constraints
  corresponding to three $SU(2)$ rotations, three spatial
  diffeomorphism and one to evolution in a timelike direction.
  Inverse of the coefficient of Nieh-Yan term, identified as
  Barbero-Immirzi parameter, acts as the coupling constant of the
  gauge theory.

\end{abstract}

\pacs{04.20.Fy, 04.60.-m, 04.60.Ds, 04.60.Pp}

\maketitle


\section{Introduction}
Addition of total divergence terms to the Lagrangian density does not 
change the classical dynamics described by it; 
  the Euler-Lagrange equations of motion are unaltered.
   In the Hamiltonian formulation, these
total divergences    reflect themselves as
canonical transformations, resulting in the change of the phase space.
This changes the symplectic structure and  Hamiltonian of the
system, yet the Hamilton's equations of motion remain equivalent to 
the Euler-Lagrange equations of the   Lagrangian formulation.

While the classical dynamics is not sensitive to the total divergence
terms in the Lagrangian density, the quantum theory may depend on
these. The canonical transformation of  classical Hamiltonian
formulation are implemented in the quantum theory through unitary
operators on the phase space   and the states. However, there are special
situations where we find topological obstructions in such a unitary
implementation. In such cases these total 
divergences    do affect the
quantum dynamics. Therefore, to have non-trivial
implications in the quantum theory, the total divergence terms have to be topological densities. This
is a necessary requirement, but not sufficient.

There are several known examples of topological terms  which have serious import in
 the quantum theory.  A well-known case is the Sine-Gordon
quantum mechanical model \cite{coleman} where an appropriate effective  topological term 
can be added to the Lagrangian density to reflect the non-perturbative properties of the quantum theory. In this model, 
we have a periodic potential with
infinitely many degenerate classical ground states. With each of 
these,  we associate a perturbative vacuum state labeled by an integer $n$
related to the {\it winding number} of homotopy maps $S^1_{} \rightarrow S^1_{}$
characterized by the homotopy group $\Pi_1^{}(S^1_{})$ which is the set of integers
${\mathbb Z}$. The {\it physical quantum vacuum state}, so called $\th${\it -vacuum},  is
non-perturbative in nature and is given by a linear superposition of
these perturbative vacua with weights given by phases $\exp({in\th})$
where angular variable $\theta$, properly normalized,  is the
coefficient of the effective topological density term in the Lagrangian
density. The physical quantities in the quantum theory depend on this
parameter. For example, the quantum vacuum energy, besides
the usual zero-point energy, has a contribution due to quantum
tunneling processes between various perturbative vacua, which depends
on $\theta$.
  
In field theory,  we  have an  example of such a topological
parameter $\theta$ in the theory of strong interactions, namely QCD.
Here also, we have infinitely many degenerate classical ground states 
labeled by integers $n$ associated with  the winding numbers of homotopy maps 
$S^3_{}\rightarrow S^3_{}$ characterized by the homotopy group 
$\Pi_3^{}(SU(3)) \equiv {\mathbb Z}$. The quantum vacuum ($\th${\it -vacuum}) is
a linear superposition of the perturbative vacua associated with these classical 
ground states. Associated effective topological   term in the Lagrangian density is
  Pontryagin density of   $SU(3)$ gauge theory with coefficient
$\theta$. This leads to  $\th$ dependent CP violating contributions to various
physical quantities.  However, there are
stringent phenomenological constraints on the value of $\th$.
 For example, from possible CP-violating contribution to
the electric-dipole moment of the neutron, this
parameter   is constrained by experimental results to be less than $10^{-10}$ radians.

In   gravity  theory in 3+1 dimensions, there are three possible
topological terms that can be added to the Lagrangian density.
Two of these, the Nieh-Yan and  Pontryagin densities, are P and T odd, and the third,
 Euler density, is P and T even. Associated
with these are three topological parameters.  In order to understand their possible import  in
the quantum theory, it is important to set up a classical Hamiltonian
formulation of the theory containing all these terms in the action.
In ref.\cite{date}, such an analysis has been presented  for a theory based 
on Lagrangian density containing  the standard Hilbert-Palatini term and  
the Nieh-Yan density \cite{nieh}.  The resulting theory,
in time gauge, has been shown to correspond to the well-known
  canonical  gauge theoretic   formulation  
  of gravity based on Sen-Ashtekar-Barbero-Immirzi    {\it real}  $SU(2)$
  gauge fields  \cite{rov}. Here  inverse of the coefficient of  
Nieh-Yan term is identified with the Barbero-Immirzi parameter
$\gamma$. Thus  the  analysis of ref.\cite{date} has  provided  a clear topological interpretation   
for $\g$,  realizing  a  suggestion   made earlier in \cite{gop} 
that  this parameter should have a topological origin.

The framework of \cite{date} involving Nieh-Yan density supersedes the earlier formulation of
Holst \cite{holst}. Detailed  Hamiltonian analysis of the theory with
Holst term for pure gravity is provided in ref.\cite{sa} and that including 
spin $1/2$ fermions in ref.\cite{mercuri}. This discussion has also been
extended to supergravity theories \cite{kaul}. Since Holst term is not
topological, {\it inclusion of matter
  necessitates matter dependent modification of the Holst term}
so that original equations of motion stay unaltered.  On the other
hand, the analysis  containing Nieh-Yan density \cite{date}, besides
explaining the topological origin of the Barbero-Immirzi parameter,
provides a {\it universal} prescription for inclusion of  
arbitrary matter without any need for
further modifications of the topological Nieh-Yan term which is given in terms
of the geometric quantities only.  As   elucidations
of these facts, this analysis has been extended to  the theory including Dirac fermions in
ref.\cite{date} and to supergravity theories in ref.\cite{sengupta}. 
 
In a quantum framework, the implications of a topological term in the
Lagrangian can also be understood through a rescaling of the wave
functional by a topologically non-trivial phase factor. This procedure
has been used for QCD \cite{jackiw} where, as mentioned above, the
properties of the non-perturbative $\th$-vacuum are effectively
represented by a $SU(3)$ Pontryagin density term in the Lagrangian.
The rescaling of wave functional is provided by the exponential of
$SU(3)$ Chern-Simons three-form with $i\th$ as its coefficient. This
framework can be extended to the gravity theory where we have a
corresponding wave functional scaling associated with the Nieh-Yan
density. However, for the pure gravity (without any matter couplings),
the standard Dirac quantization, where the second class constraints
are implemented before quantization, is not appropriate. This is so
because second class constraints of pure gravity imply vanishing of
the torsion, which results in making the rescaling trivial.  Instead,
as discussed in \cite{sengupta1}, the Gupta-Bleuler and coherent state
quantization methods are well suited for the purpose. These methods
are quite general and can be used for gravity theory with or without
matter. However, for matter-couplings leading to non-vanishing
torsion, {\it e.g.} Dirac fermions, the Dirac quantization, as has
been discussed earlier in ref.\cite{mercuri2}, can also be adopted for
this purpose.

Hamiltonian analysis of the first order (anti-) self-dual Lagrangian
density for gravity including the Pontryagin density of {\it complex}
$SU(2)$ (anti-) self-dual gauge fields has first been reported by
Montesinos in \cite{mon}. In the time gauge, the Sen-Ashtekar {\it
  complex} $SU(2)$ connection stays unchanged, but its conjugate
momentum field gets modified by the presence of the Pontryagin term.
Recently, in \cite{rp}, this analysis has also been done for gravity
theories containing Holst, Nieh-Yan, Euler and Pontryagin terms. This
study concludes that, in the time gauge, {\it real} $SU(2)$ gauge
theoretical formulation is possible {\it only if} the Pontryagin and
Euler terms are absent; the Pontryagin density can be added consistently only in the
{\it complex} $SU(2)$ gauge formulation leading to a canonical
analysis in accordance with results of Montesinos \cite{mon}.
 
In the following, we present a classical Hamiltonian analysis for
theory of gravity based on Hilbert-Palatini Lagrangian supplemented
with {\it all the three possible topological terms in $(1+3)$
  dimensions}, namely, Nieh-Yan, Pontryagin and Euler classes.  Unlike
\cite{rp}, in view of results of \cite{date} and the remarks already
made above, we shall not add the Holst term, which is not a
topological density.  We demonstrate that, in the time gauge, we do
have {\it a real $SU(2)$ gauge theory} with its coupling given by
inverse of the coefficient of Nieh-Yan term. The canonical theory also
depends on {\it two additional arbitrary parameters}, the coefficients
of Pontryagin and Euler terms in the Lagrangian density. These
parameters are not subjected to any restrictions.  A formulation of
the theory presented involves the standard
Sen-Ashtekar-Barbero-Immirzi real $SU(2)$ connections $A^i_a$, which
depend only on the coefficient of the Nieh-Yan term, as the canonical
fields. Associated conjugate momentum fields, instead of being
densitized triads of the standard canonical theory, are modified and
depend on the coefficients of the Nieh-Yan, Pontryagin and Euler
terms.  There are second class constraints in the description,
essentially reflecting the fact that the extrinsic curvature is not
independent.  Correspondingly, for this constrained Hamiltonian
system, the Dirac brackets analysis is developed. Dirac brackets of
the phase variables do not exhibit the same algebraic structure as
those of the standard canonical theory of gauge fields $A^i_a$ and
densitized triads $E^a_i$; the new variables are not related to them
by a canonical transformation. However, it is possible to construct
another set of phase variables which are canonical transforms of the
standard variables $(A^i_a,~ E^a_i)$.  In this framework, both new
gauge fields and their conjugate momentum fields are modified and
develop dependences on all three topological parameters. The canonical
formulation described in terms these new phase variables is presented
in detail.

\section{Topological coupling constants in gravity}

 We set up the standard  theory of pure ({\it i.e.}, no matter couplings)
gravity in terms of the $24$ ~$SO(1,3)$ gauge 
connections $\w_\m^{IJ}$ and $16$ tetrad fields $e^I_\m$ as the 
{\it independent}
fields described by 
Hilbert-Palatini (HP) Lagrangian density:
\begin{eqnarray}
{\cal L}^{}_{HP} ~=~ {\frac 1 2}~ e~
\Sigma^{\m\n}_{IJ} ~ R^{~~~IJ}_{\m\n}(\w) \label{HP}
\end{eqnarray}
where 
\beq
e ~\equiv ~ det(e_\m^I)~, ~~~~ &&\Sigma^{\m\n}_{IJ} ~\equiv~ {\frac 1 2}~ e^\m_{[I} e^\n_{J]} 
~\equiv ~ {\frac 1 2} ~\left( e^\m_I e^\n_J - e^\m_J e^\n_I \right)~, ~\nonumber \\
R^{~~~IJ}_{\m\n}(\w) &\equiv &\partial_{[\m} \w_{\n]}^{~IJ}
+\w_{[\m}^{~IK} \w_{\n]K}^{~~~~J} 
\eeq
and $e^\m_I$ is the inverse of the tetrad field, $e^\m_I ~e^I_\n~= \delta^\m_{~\n}$~, $~~ e^I_\m ~e^\m_J~=~ \delta^I_{~J}$.

Modifications of the gravity  Lagrangian density  by terms which are quadratic in curvature and particularly also
include torsion, without  altering  the field equations, have a long history, see for example \cite{hehl}.

In $(1+3)$ dimensions, there are three possible topological terms
that can be added  to  the HP Lagrangian density (\ref{HP}). These are:

{\it (i) Nieh-Yan class:}\cite{nieh} 
\beq
I_{NY}^{} ~= ~e\Sigma^{\m\n}_{IJ}
{\tilde R}^{~~~IJ}_{\m\n}(\w) ~+ ~\e^{\m\n\a\b}
D_\m(\w)e_{I\n} D_\a(\w)e^I_\b  \label{NY}
\eeq
where the dual in the internal space is defined as:
\beq
{\tilde X}^{IJ}_{}  ~\equiv~
{\frac 1 2} ~\e^{IJKL} ~X_{ KL}^{} \nonumber
\eeq
and the $SO(1,3)$ covariant derivative is: $D^{}_\m (\w) e^I_\n = \partial^{}_\m e^I_\n + \w^{~~I}_{\m~~J} e^J_{}$. 

This topological density involves torsion. It can be explicitly written as a total divergence as:  
\beq 
I_{NY}^{} \equiv~  \partial_\m
\left[ \e^{\m\n\a\b} ~e^I_\n~ D_\a(\w) e_{I\b} \right]  \label{NY2}
\eeq
In the Euclidean theory, as discussed in \cite{cz},
 this topological 
density, properly normalized,  characterizes the {\it winding numbers} given by three integers
associated with the homotopy groups $\Pi_3(SO(5)) = {\mathbb Z}$ and $\Pi_3(SO(4)) = ({\mathbb Z}, ~{\mathbb Z})$.

{\it (ii)  Pontryagin class:}
\beq
I_P^{} 
~= ~\e^{\m\n\a\b} R_{\m\n IJ} (\w) R_{\a\b}^{~~~IJ}(\w)  \label{P}
\eeq
This is  the same topological density as   in the case of QCD except that
the gauge group here is $SO(1,3)$ instead of $SU(3)$. Again, it is a total divergence, given in terms of the
$SO(1,3)$ Chern-Simons three-form:
\beq
I^{}_P ~\equiv ~4\partial_\m
\left[\e^{\m\n\a\b} \w^{~IJ}_\n \left( \partial_\a \w_{\b IJ}
+ {\frac 2 3}  \w^{~~~K}_{\a I} \w_{\b KJ} \right)\right] \label{Po2}
\eeq
 For the Euclidean theory, this topological density,
properly normalized, characterizes the {\it winding numbers} given by two integers corresponding
to the homotopy group $\Pi_3(SO(4)) = ({\mathbb Z}, ~ {\mathbb Z})$.

{\it (iii) Euler class:} \beq I_E^{} ~=~ \e^{\m\n\a\b} R_{\m\n IJ}
(\w) {\tilde R}_{\a\b}^{~~~IJ}(\w) \label{E} \eeq which again is a
total divergence which can be explicitly written as: \beq I^{}_E
~\equiv~ 4\partial_\m \left[\e^{\m\n\a\b} {\tilde\w}^{~IJ}_\n
  \left(\partial_\a \w_{\b IJ} + {\frac 2 3} \w^{~~~K}_{\a I} \w_{\b
      KJ} \right)\right] \label{Eu2} \eeq For the Euclidean theory,
integral of this topological density, properly normalized, over a
compact four-manifold is an alternating sum of Betti numbers 
$b^{}_0 -b^{}_1 + b^{}_2 - b^{}_3 $ , characterizing the manifold.

Now we may construct the most general Lagrangian density by adding these topological terms
(\ref{NY}), (\ref{P}) and (\ref{E}), with the coefficients $\eta$, $\theta$ and $\phi$ 
respectively, to the Hilbert-Palatini Lagrangian density (\ref{HP}).
Since all the topological terms are
total divergences, the classical equations of motion are independent of 
the  parameters $\eta$, $ \th$ and $\phi$.
However, the Hamiltonian formulation  and the symplectic structure 
do see these parameters. Yet, classical  dynamics are
independent of them. But, quantum theory may depend on them. 
 
All these topological terms in the action  are  functionals of local  geometric quantities,  yet they
represent only the topological properties of the four-manifolds. These do not change under 
continous deformations of the  four-manifold geometry.

Notice that, while  the Nieh-Yan $ I^{}_{NY}$ and Pontryagin $I^{}_P$ densities are P and T violating,  
the Euler density $I^{}_E$ is not. 
So in  a  quantum theory of gravity including these terms, besides the Newton's coupling constant,
we can have three  additional  dimensionless
 coupling constants, two P and T  violating ($\eta$,~ $\th$) and one P and T  preserving ($\phi$).

 \section{Hamiltonian formulation of gravity with Nieh-Yan, Pontryagin  and Euler densities}

 Here we shall carry out the Hamiltonian analysis for the most general
 Lagrangian density containing all   three topological terms besides the Hilbert-Palatini term: 
 \beq {\cal L}
 ~=~ {\frac 1 2}~ e~ \Sigma^{\m\n}_{IJ} ~ R^{~~~IJ}_{\m\n}(\w) ~+~
 {\frac \eta 2} ~I^{}_{NY} ~+~ {\frac \th 4} ~I^{}_P ~+ ~{\frac \phi
   4} ~I^{}_E   \label{L1} \eeq
 where the Nieh-Yan $I_{NY}^{}$, Pontryagin $I_P^{}$ and Euler $I_E^{}$ densities  are given by (\ref{NY}), 
 (\ref{P}) and (\ref{E}) respectively.

We shall use the following parametrization for tetrad fields\footnote{This parametrization differs from the 
one  used earlier in \cite{date}. To obtain the present parametrization replace $eN$ by $N^2_{}$ in the earlier parametrization.}: 
\beq
e^{I}_{t} & =& N M^{I}+N^{a}V_{a}^{I}, ~~~~~ e^{I}_{a} =
V^{I}_{a}~; \nonumber \\
M_{I}V_{a}^{I} & = & 0, ~~~~~ M_{I}M^{I} = -1  \label{tetpara1}
\eeq
 with $N$ and $N^a_{}$ as the lapse and shift fields. The inverse tetrads are:
\beq e^{t}_{I} & =& -\frac{M_{I}}{N}~, ~~~~~ e^{a}_{I} =
V^{a}_{I}+\frac{N^{a}M_{I}}{N}; \nonumber \\
M^{I}V_{I}^{a} &=& 0~,~~~~ V_a^I V^b_I = \delta_a^b~, ~~~~~~~ V_a^I
V^a_J = \delta^I_J + M^IM_J  \label{tetpara2}
 \eeq 
The internal space metric is $\eta^{IJ}_{} \equiv dia(-1,~1,~1,~1)$. The three-space metric is $q_{ab} \equiv 
V^I_a V^{}_{bI}$ with $q = det (q^{}_{ab})$ which leads to $e \equiv det (e^I_\m)$ $= N{\sqrt q}$. The inverse 
three-space metric is 
$q^{ab} = V^a_I V^{bI}_{}$, ~ $ q^{ab}_{} q_{bc}^{} = \delta^a_c$.  Two useful identities are:
\beq
2e\S^{ta}_{IJ} ~=~ -{\sqrt q} ~ M^{}_{[I}V^a_{J]} ~,
~~~~~~~~e \S^{ab}_{IJ} = \frac{2Ne^2_{}}{{\sqrt q}~} \S^{t[a}_{IK} \S^{b]t}_{JL} \eta^{KL}_{} 
+ e~ N^{[a}_{} \S^{b]t}_{IJ}
\label{identity1}\eeq

In this parametrization, we have, instead of the $16$ tetrad components $e^I_\m$, the following
$16$ fields: 9 $V^{a}_{I}$ $(M^{I}V^{a}_{I}= 0)$, 3 $M^I$
$(M^I_{}M^{}_I =- 1)$ and 4 lapse and shift vector fields $N$,
$N^{a}$. From these, instead of the  variables $V^{a}_{I}$ and $M^{I}$, we define a convenient set of 12 variables,
 as: 
\beq
~E^{a}_{i} &=& 2e\Sigma^{ta}_{0i}~\equiv~ e\left(e^t_0 e^a_i - e^t_i
  e^a_0\right)= - ~ {\sqrt q}~ M^{}_{[0} V^a_{i]}, ~~~~~~~~\chi_{i} = -M_{i}/M^{0}~ ~~~~ \label{E1}
   \eeq
which further imply:
\beq
2e\S^{ta}_{ij} = - ~{\sqrt q} ~ M^{}_{[i} V^a_{j]} = - E^a_{[i} \chi^{}_{j]} \label{E2}
\eeq   
  
 Now, using the parametrization (\ref{tetpara1}, \ref{tetpara2}) for the tetrads, and the second identity in (\ref{identity1}), we expand the various terms to write:
 \beq
   {\frac 1 2}~ e ~\Sigma^{\m\n}_{IJ} ~
   R^{~~~IJ}_{\m\n}(\w) + {\frac \eta 2} ~I^{}_{NY}
   ~=~e\Sigma^{ta}_{IJ}\del_{t}\omega_{a}^{(\eta)IJ}+t^{a}_{I}\del_{t}V_{a}^{I}-NH-N^{a}H_{a}-\frac{1}{2}\omega_{t}^{IJ}G_{IJ} \label{L2}
\eeq
where we have dropped the total space derivative terms. Here  $ t^{a}_{I}\equiv\eta \epsilon^{abc}D_{b}(\omega)V_{cI}$ 
with $\e^{abc}_{} \equiv \e^{tabc}_{}$ and, for any internal space antisymmetric tensor, $X^{(\eta)}_{IJ}  \equiv X_{IJ}^{}  + \eta  {\tilde X}^{}_{IJ}$
$=X^{}_{IJ} + \frac{\eta}{2} \e^{}_{IJKL}  X_{}^{KL}$. Further, 
\beq
  H~&=&~\frac{2e^2}{{\sqrt q}~} \Sigma^{ta}_{IK}\Sigma^{tb}_{JL}\eta^{KL}R_{ab}^{ IJ}(\omega)  
   ~= ~\frac{2e^2}{{\sqrt q}~} \Sigma^{ta}_{IK}\Sigma^{tb}_{JL}\eta^{KL}R_{ab}^{(\eta)IJ}(\omega)-M^{I}D_{a}(\omega)t^{a}_{I} ~~~~ \nonumber \label{H1}\\
  H_{a}~&=&~e\Sigma^{tb}_{IJ}R_{ab}^{IJ}(\w)   ~=~ 
  e\Sigma^{tb}_{IJ}R_{ab}^{(\eta)IJ}(\w)~-~V_{a}^{I}D_{b}(\omega)t^{b}_{I} \nonumber \label{Ha1}\\
  G_{IJ}~&=&~ -2 D_{a}(\w)\left\{e\Sigma^{ ta}_{IJ}\right\}~=~ -2D_{a}(\w)\left\{e\Sigma^{(\eta)ta}_{IJ}\right\} -t^a_{[I}V^{}_{J]a}\label{G1}
\eeq
where   we have used the following identities:
\beq
M^I_{} D^{}_a(\w)t^a_I  &\equiv &  \frac{2\eta e^2_{}}{\sqrt q} \S^{ta}_{IK} \S^{tb}_{JL} \eta^{KL}{} {\tilde R}^{~~IJ}_{ab}(\w)  \nonumber \\
V^I_a D^{}_b(\w) t^b_I  & \equiv&  \eta e\S^{tb}_{IJ} {\tilde R}^{~~IJ}_{ab}(\w)~, ~~~~~~~~~~ 
t^a_{[I}V^{}_{J]a}  \equiv   - 2 \eta D^{}_a(\w)\left\{e {\tilde {\S}}^{ta}_{IJ} \right\}\nonumber
\eeq

Next notice that, dropping the total space derivative terms and using the Bianchi identity, 
$\e^{abc}_{} D^{}_a(\w) R^{}_{bcIJ} \equiv 0$, we can write
\beq
\frac{\th}{4} ~I_P^{}~ + ~\frac{\phi}{4} ~I_E^{} ~=  ~e^a_{IJ} ~ \partial_t^{} \w^{(\eta)IJ}_a \label{L3}
\eeq
where   $e^a_{IJ}$ are given by
\beq
 \left(1+\eta^2_{}\right) e^a_{IJ} ~ = ~ \e^{abc}_{}
\left\{ \left(\th+\eta \phi\right) R^{}_{bcIJ}(\w) + \left( \phi - \eta \th \right) {\tilde R}^{}_{bcIJ}(\w) \right\} 
\label{eIJ} 
\eeq

Thus, collecting terms from (\ref{L2}) and (\ref{L3}), full Lagrangian density (\ref{L1}) assumes the following form:
\beq
 {\cal L}~=~\pi^{a}_{IJ}\del_{t}\omega_{a}^{(\eta)IJ}~+~t^{a}_{I}\del_{t}V_{a}^{I}~-~NH~-~N^{a}H_{a}~-~\frac{1}{2}\omega_{t}^{IJ}G_{IJ} \label{L4}
\eeq
with
\beq
\pi^{a}_{IJ}~=~e~\Sigma^{ta}_{IJ}~+~e^{a}_{IJ}
\eeq
In this Lagrangian density, the fields $\w^{(\eta)IJ}_a$ and $\pi^a_{IJ}$ form   canonical pairs. Then, $H$, $H^{}_a$ and $G^{}_{IJ}$ of (\ref{H1})  can be expressed in terms of these fields as:
\beq
G_{IJ}~&=&~   -2D_{a}(\w)\pi^{a(\eta) }_{IJ} -t^a_{[I}V^{}_{J]a}\label{G2} \\
H_{a}~&=&~ 
  \pi^{b}_{IJ}R_{ab}^{(\eta)IJ}(\w)~-~V_{a}^{I}D_{b}(\omega)t^{b}_{I} \label{Ha2}\\
 H~&=&~\frac{2 }{{\sqrt q}~} \left( \pi^{a(\eta) }_{IK} -e^{a(\eta) } _{IK}\right)\left(\pi^{b(\eta) }_{JL}
 -e^{b(\eta)}_{JL}\right) \eta^{KL}R_{ab}^{~~ IJ}(\omega) -M^{I}D_{a}(\omega)t^{a}_{I} ~~~~ \label{H2}  
   \eeq
  where we have used the relations: $ D^{}_a(\w)e^a_{IJ}  =0$  and $ D^{}_a(\w){\tilde e}^a_{IJ}  =0$ which result from   the Bianchi identity
  $\e^{abc}_{} D^{}_a(\w) R^{}_{bcIJ}(\w) =0$,  and also used $e^b_{IJ} R^{~~IJ}_{ab}(\w) =0$ and ${\tilde e}^b_{IJ} R^{~~IJ}_{ab} (\w) =0$ which follow from the fact that 
  $2q\left( \th^2_{} + \phi^2_{} \right) R^{}_{abIJ} = \e^{}_{abc}\left\{ \left( \th + \eta \phi \right) e^c_{IJ} -
  \left( \phi - \eta \th \right) {\tilde e}^c_{IJ} \right\}$.

Now, in order to unravel the $SU(2)$ gauge theoretic framework for the  Hamiltonian formulation, from the 24 $SO(1,3)$ gauge fields $\w^{~IJ}_\m$, we define, in addition to 6 field variables $\w^{IJ}_t$,
  the following   suitable set of 18 field variables: 
\beq A^i_a~\equiv~ \w^{(\eta)0i}_a =\w^{0i}_a
  +\eta{\tilde {\w}}^{0i}_a, ~~~~~~~K^i_a ~\equiv~ \w^{0i}_a
\eeq
The fields  $A^i_a$   transform as the  connection
 and the extrinsic curvature $K^i_a$ as   adjoint representations  under the $SU(2)$ gauge transformations.
In terms of these, it is straight forward to check that:
 \beq
\pi^{a}_{IJ}\del_{t}\omega_{a}^{(\eta)IJ}~ =~2\pi^{a}_{0i}\del_{t}\omega_{a}^{(\eta)0i}
~+~\pi^{a}_{ij}\del_{t}\omega_{a}^{(\eta)ij}  
~&=&~\hat{E}^{a}_{i}\del_{t}A_{a}^{i}~+~\hat{F}^{a}_{i}\del_{t}K_{a}^{i} \label{L5}
\eeq
 with
 \begin{eqnarray} 
 {\hat E}^i_a ~&\equiv&~-~\frac{2}{\eta}
  ~\tilde{\pi}^{a(\eta) }_{0i}~\equiv~ -~\frac{2}{\eta}
 ~ \left( {\tilde \pi}^a_{0i}- \eta \pi^a_{0i} \right) ~=~ E^a_i   -\frac{2}{\eta}~{\tilde e}^{a(\eta) }_{0i}(A,K) +\frac{1}{\eta}~\e^{ijk}E^a_j\chi^{}_k 
  \\
  {\hat F}^a_i ~&\equiv&~2\left(\eta
    +\frac{1}{\eta}\right)\tilde{\pi}^{a}_{0i}~=~ \left( \eta +
    \frac{1}{\eta} \right) \left\{ -\epsilon^{ijk}E^{a}_{j}\chi_{k}+ 2
  {\tilde e}^a_{0i}(A,K)\right\}\label{F}
\end{eqnarray}
where  $e^a_{0i}$ and  ${\tilde e}^a_{oi} \equiv \frac{1}{2} \e^{ijk}_{} e^a_{jk}$ as defined in (\ref{eIJ}) and
${\tilde e}^{a(\eta) }_{0i} \equiv {\tilde e}^a_{0i} - \eta e^a_{0i}$   are  written as   functions of  the gauge field $A^i_a$   and  the extrinsic curvature $K_a^i$ using
\beq
R^{~~0i}_{ab}(\w) &= &D^{}_{[a} (A) K^i_{b]} ~-~ \frac{2}{\eta} ~\e^{ijk}_{} K^j_a K^j_b  \nonumber  \\
R^{~~ij}_{ab} (\w) &=& - ~\frac{1}{\eta} ~\e^{ijk}_{} F^k_{ab}(A) ~+ ~\frac{1}{\eta} ~\e^{ijk}_{}D^{}_{[a} (A) K^k_{b]}
~-\left(\frac{\eta^2_{} -1}{\eta^2{}}\right)  K^i_{[a} K^j_{b]} \label{R}
\eeq
with the $SU(2)$ field strength and covariant derivative respectively as:
\beq
F^i_{ab} (A) \equiv   \partial^{}_{[a}  A^i_{b]} ~+~ \frac{1}{\eta} ~\e^{ijk}_{} A^j_a A^k_b ~, ~~~~
D^{}_a(A) K^i_b \equiv   \partial^{}_a K^i_b ~+ ~\frac{1}{\eta} ~\e^{ijk}_{} A^j_a K^k_b
\eeq

Now, using (\ref{L5}), the Lagrangian density (\ref{L4}) can be written as:
\beq
{\cal L}~=~~\hat{E}^{a}_{i}\del_{t}A_{a}^{i}~+~\hat{F}^{a}_{i}\del_{t}K_{a}^{i}~+~t^{a}_{I}\del_{t}V_{a}^{I}~-~NH~-~N^{a}H_{a}~-~\frac{1}{2}~\omega_{t}^{IJ}G_{IJ} \label{L6}
\eeq
 Thus, we have the canonically conjugate pairs $(A^i_a, ~ {\hat
  E}^a_i)$, $(K^i_a,~ {\hat F}^a_i)$ and $(V^I_a,~ t^a_I)$. We may write $G^{}_{IJ}$, $H_a^{}$ and $H$ of
  (\ref{G2})-(\ref{H2}) in terms of 
  these fields. For example, from (\ref{G2}):
  \beq
  G^{rot}_i &\equiv &\frac{1}{2} ~\e_{ijk}^{} G_{jk} = \eta D^{}_a(A){\hat E}^a_i +~\e^{}_{ijk} \left( K^j_a {\hat F}^a_k 
  -t^a_jV^k_a\right)   \label{rotconstraints}\\
  G^{boost}_i &\equiv & G^{}_{0i} = - D^{}_a(A) \left( {\hat E}^a_i + {\hat F}^a_i\right)
  ~+~ \e^{ijk}_{} K^j_a \left\{ \left( \eta + \frac{1}{\eta} \right) {\hat E}^a_k + \frac{1}{\eta} {\hat F}^a_k \right\}
  - t^a_{[0} V^{}_{i]a}  \label{boostconstraints} \\
  &&~~~~= -D^{}_a(A) {\hat F}^a_i + \e^{ijk}_{} K^j_a\left\{ \left( \eta+ \frac{1}{\eta} \right) {\hat E}^a_k + \frac{2}{\eta} {\hat F}^a_k \right\} -\frac{1}{\eta} \e^{ijk}_{} t^a_j V^{}_{ak} - t^a_{[0} V^{}_{i]a}   -\frac{1}{\eta} ~G^{rot}_i
  \nonumber 
  \eeq
  where the covariant derivatives are: 
 $  D^{}_a(A){\hat E}^b_i  ~=~ \partial^{}_a{\hat E}^b_i ~+~\eta^{-1}_{} ~\e^{ijk}_{}A^j_a {\hat E}^b_k$ and 
    $ D^{}_a(A){\hat F}^b_i  =~ \partial^{}_a{\hat F}^b_i  $ $ ~+~ \eta^{-1}_{} ~\e^{ijk}_{}A^j_a {\hat F}^b_k $.  
 Next,  for the generators of spatial diffeomorphisms $H_a$ from (\ref{Ha2}):
 \beq
  H^{}_a &= &{\hat E}^b_i F^i_{ab}(A) + {\hat F}^b_i D^{}_{[a}(A)
K^i_{b]} - K^i_a D^{}_b(A) {\hat F}^b_i +
t^b_i D^{}_{[a}(A) V^i_{b]} - V^i_a D^{}_b(A)t^b_i \nonumber \\
&&~~~~~~~~~~~~+t^b_0\partial_{[a} V^0_{b]}-V^0_a \partial_b^{}t^b_0 - \frac{1}{\eta} \left( G^{rot}_i +\eta G^{boost}_i\right)K^i_a  \nonumber \\
&=& {\hat E}^b_i \partial^{}_{[a} A^i_{b]} - A^i_a
\partial^{}_b{\hat E}^b_i + {\hat F}^b_i \partial^{}_{[a} K^i_{b]} -
K^i_a\partial^{}_b{\hat F}^b_i
+t^b_i\partial^{}_{[a} V^i_{b]} - V^i_a\partial^{}_b t^b_i \nonumber \\
&&~~~~~~~~~~~~+t^b_0\partial^{}_{[a}V^0_{b]} - V^0_a \partial^{}_b t^b_0+ \frac{1}{\eta} G^{rot}_i  A^i_a 
-\frac{1}{\eta} \left(G^{rot}_i + \eta G^{boost}_i\right)K^i_a 
   \eeq
   where we have used
   $
   -V^I_a D^{}_b(\w)t^b_I \equiv - V^I \partial^{}_b t^b_I + t^b_I \partial^{}_{[a}V^I_{b]} + t^b_I V^{}_{Jb} \w^{IJ}_a  $. ~
 Similarly we can express $H$ of (\ref{H2}) in terms of these fields.
 
  Now, notice that all the  fields $(A^i_a, ~ {\hat
  E}^a_i)$, $(K^i_a,~ {\hat F}^a_i)$ and $(V^I_a,~ t^a_I)$ in the Lagrangian density (\ref{L6})
  are not independent. Of these, the fields $V^I_a$ and $t^a_I$ are given in terms of others as:
  $V^I_a =v^I_a$ and $t^a_I =\tau^a_I$  with
 \beq 
 v^i_a ~\equiv ~\frac{1}{\sqrt E} E^i_a ~, ~~~~~~~~~~v^0_a \equiv -\frac{1}{\sqrt E} E^i_a \chi_i^{} \label{v}
\eeq
where   $E_a^i$ is inverse of  
$E^a_i$, {\it i.e.}, $ E^i_aE^b_i = \delta^b_a$, ~$ E^i_a E^a_j = \delta^i_j$  and $E~\equiv~ det (E^i_a) =q^{-1}(M^0_{})^{-2}$ and 
\beq 
\tau^a_i &\equiv & \eta \e^{abc}_{} D_b(\w) v^i_c ~= ~ \e^{abc}_{} \left( \eta D_b(A)
  v^i_c - \e^{ijk}_{} K^j_b v^k_c + K^i_b v^0_c\right) ~, \nonumber \\
  \tau^a_0 &\equiv& -\eta \e^{abc}_{}D^{}_b(\w)v^0_c = -\eta \e^{abc}_{} \left( \partial^{}_b v^0_c + K^j_bv^j_c \right)
  \label{tau1}
   \eeq 
    In addition, the fields $\hat {F}^{a}_{i}$, which are conjugate to the extrinsic  curvature $K^i_a$, are also not independent; these are given  in terms  of other   fields by (\ref{F}) .
    
 In
 the Lagrangian density (\ref{L6}), there are no velocity terms associated with 
 $SO(1,3)$ gauge fields $\w^{IJ}_t$,   shift vector field $N^{}_a$ and  lapse field $N$.  Hence these fields 
are Lagrange multipliers. Associated with these  are as many constraints: ~$G^{}_{IJ} \approx 0$, ~$H^{}_a \approx 0$,~ 
 and $H\approx 0$ where the weak equality $\approx$ is in the sense of Dirac theory of constrained Hamiltonian systems.
 Here from the form of $G^{rot}_i = \frac{1}{2}~\e^{}_{ijk} G^{}_{jk}$  in (\ref{rotconstraints}), it is clear that these generate $SU(2)$ rotations on various fields. The boost transformations 
 are generated by $G^{boost}_i = G^{}_{0i}$,  spatial diffeomorphisms by $H^{}_a $   and $H \approx 0$ 
 is the   Hamiltonian constraint. This, thus can already be viewed,  {\it without fixing the boost degrees of freedom
 and without solving the second class constraints (\ref{v}) and (\ref{tau1})},  as a $SU(2)$ gauge 
 theoretic framework. Here,  besides the three  $SU(2)$ generators $G^{rot}_i$, we  have seven  
  constraints, $G^{boost}_i, ~ H^{}_a$ and $ H$. 
 We may, however, fix the boost gauge invariance  by choosing a time gauge.
  Then we are left with only the $SU(2)$ gauge invariance besides the diffeomorphism 
  $H_a^{}$  and Hamiltonian   $H$  constraints.
  This we do in the next section.
  
\section{Time gauge} 
We  work in the time (boost) gauge by choosing the gauge condition $\chi_i^{} =0$ which then implies for
the tetrad components   $  e^0_a   \equiv V^0_a   =0$.  Correspondingly the  boost generators 
(\ref{boostconstraints}) are also   set equal to zero 
strongly, ~$G^{boost}_i =0$. 
 In this gauge, the Lagrangian density (\ref{L6}) takes the simple form:
 \beq
{\cal L}=~\hat{E}^{a}_{i}\del_{t}A_{a}^{i}~+~\hat{F}^{a}_{i}\del_{t}K_{a}^{i}~+~t^{a}_{i}\del_{t}V_{a}^i - {\cal H} 
\label{L7}
\eeq
with the Hamiltonian density as:
\beq
{\cal H} ~=~NH~+~N^{a}H_{a}~~+~
{\frac 1 2} ~\e^{ijk}_{} \omega_{t}^{ij}G^{rot}_{k} ~+~ \xi^a_i\left( V^i_a - v^i_a\right)  \nonumber \\
  +~\phi^i_a\left( t^a_i - \tau^a_i\right) ~+~ \lambda^i_a \left\{
  {\hat F}_i^a - 2 \left( \eta + \frac{1}{\eta}\right) {\tilde
    e}^a_{0i} (A,K)\right\}  \label{Ham1} 
\eeq
 where all the fields involved are not independent. In particular, the fields $V^i_a$, $t^a_i$ and ${\hat F}^a_i$ depend 
on other fields. This fact is reflected in   ${\cal H}$ above
 through   terms with   Lagrange multiplier fields $\xi^a_i$, $\phi^i_a$ and
$\lambda^i_a$. Now, in this time gauge, expressions for $G^{rot}_i$, $H^{}_a$ and $H$ are:
 \beq
&&G^{rot}_i ~\equiv~ \eta D^{}_a(A) {\hat E}^a_i + \e^{}_{ijk} \left( K^j_a{\hat F}^a_k - t^a_jV_a^k \right) \nonumber\\
&&H^{}_a \equiv~ {\hat E}^b_i F^i_{ab}(A) + {\hat F}^b_i D^{}_{[a}(A)
K^i_{b]} - K^i_a D^{}_b(A) {\hat F}^b_i +
t^b_i D^{}_{[a}(A) V^i_{b]} - V^i_a D^{}_b(A)t^b_i  -  {\eta}^{-1}_{} G^{rot}_i K^i_a   \nonumber \\
&&~~~=~ {\hat E}^b_i \partial^{}_{[a} A^i_{b]} - A^i_a
\partial^{}_b{\hat E}^b_i + {\hat F}^b_i \partial^{}_{[a} K^i_{b]} -
K^i_a\partial^{}_b{\hat F}^b_i
+t^b_i\partial^{}_{[a} V^i_{b]} - V^i_a\partial^{}_b t^b_i +  {\eta}^{-1}_{} G^{rot}_i\left( A^i_a -K^i_a \right) 
\nonumber  \\
&&H~\equiv~ \frac{\sqrt E}{2\eta} \e^{ijk}_{} E^a_i E^b_j \left\{
  F^k_{ab}(A) -\left(1+\eta^2_{}\right)
  \left(D^{}_{[a}(A)K^k_{b]} - {\eta}^{-1}_{} ~\e^{kmn}_{} K^m_a K^n_b\right)  \right\}~ \nonumber \\
&&
~~~~~~~~~~~~~~~~~~~~~~~~~~~~~~~~~~~~~~~~~~~~~~~~~~~~~~~~~~~~~+~K^i_a
t^a_i~ -{\eta}~\partial^{}_a\left( {\sqrt E} G^{rot}_k E^a_k\right)  \label{GHHconstraints}
\eeq 
where $D_a(A)$ is the $SU(2)$ gauge covariant derivative. In
the last line, we have used the time-gauge identity: $t^a_0 =\tau^a_0= \eta {\sqrt E} G^{rot}_k E^a_k$.
Also  $E^a_i$ are  functions of ${\hat E }^a_i$, $A^i_a$ and $K^i_a$:
\beq
  E^a_i ~= ~E^a_i ({\hat E}, A, K)~\equiv~ {\hat E}^a_i ~+~
\frac{2}{\eta}~{\tilde e}^{a(\eta) }_{0i} (A,K)\label{Econd}
\eeq.
 
Associated with the Lagrange multiplier fields
    $\w^{ij}_t$, $ N^a_{}$ and $N$  in (\ref{Ham1}),  we have the constraints:
 \beq 
 G_{i}^{rot} ~\approx ~0,~~~~~~~~ H_a ~\approx ~ 0 ~, ~~~~~~~H ~\approx~ 0 
\eeq 
In addition, corresponding to Lagrange multiplier fields $\xi^a_i$ and $\phi^i_a$, we have more constraints: 
 \beq
 V^i_a ~-~ v^i_a(E) ~\approx ~0~, ~~~~~~~~~ t^a_i  ~-~ \tau^a_i(A,K,E) ~ \approx ~ 0  \label{tau2}
\eeq
where, from (\ref{v}) and (\ref{tau1}), in the time gauge:
\beq
v^i_a ~\equiv ~\frac{1}{\sqrt E}~ E^i_a ~,  ~~~~~~~~~~~
\tau^a_i  \equiv   \eta \e^{abc}_{} D_b(\w) v^i_c ~= ~ \e^{abc}_{} \left( \eta D_b(A)
  v^i_c - \e^{ijk}_{} K^j_b v^k_c  \right) ~ \label{tau3}
   \eeq 
  Similarly, from the last term in (\ref{Ham1}), there are the additional constraints:
  \beq
\chi^a_i ~\equiv ~ {\hat F}^a_i ~-~ 2 \left( \eta+ \frac{1}{\eta}
\right) {\tilde e}^a_{0i}(A,K) ~\approx~0  \label{chi}
\eeq 
Here $e^a_{0i}$
and ${\tilde e}^a_{0i}$ of (\ref{eIJ}), with the help of Eqn.(\ref{R}), are written as  functions of the gauge fields $A^i_a$,  
extrinsic curvature $K^i_a$ and the topological parameters $\theta$,
$\phi$ besides $\eta$ as follows: 
\beq 
&&\eta^2_{}\left( 1+
  \eta^2_{}\right) e^a_{0i}(A,K) \equiv  - \e^{abc}_{} \left\{\eta
  \left( \phi-\eta\theta\right) F^i_{bc}(A)
  -2 \eta \left(\left(1-\eta^2_{}\right) \phi -2\eta \theta \right) D^{}_b (A)K^i_c  \right. \nonumber \\
&&~~~~~~~~~~~~~~~~~~~~~~~~~~~~~~~~~~~~~~~~~~~~~~~~\left. - \left( \eta\left(3-\eta^2_{}\right)\theta+\left(3\eta^2_{} -1\right) \phi \right) \e^{ijk}_{} K^j_b K^k_c \right\} \nonumber \\
&& \eta^2_{}\left( 1+ \eta^2_{}\right){\tilde e}^a_{0i}(A,K) \equiv  -
\e^{abc}_{} \left\{\eta \left( \theta+\eta\phi\right) F^i_{bc}(A)
  -2 \eta \left(\left(1-\eta^2_{}\right) \theta +2\eta \phi\right) D^{}_b(A)K^i_c  \right. \nonumber \\
&&~~~~~~~~~~~~~~~~~~~~~~~~~~~~~~~~~~~~~~~~~~~~~~~\left. - \left( \left(3
      \eta^2_{}-1\right)\theta -\eta\left(3-\eta^2_{} \right) \phi
  \right) \e^{ijk}_{} K^j_b K^k_c \right\} \label{e(A,K)}
  \eeq
   From these we can construct for $e^{a(\eta)}_{0i} \equiv e^a_{0i} +\eta {\tilde e}^a_{0i}$ 
  and ${\tilde e}^{a(\eta)}_{0i} \equiv {\tilde e}^a_{0i} -\eta   e ^a_{0i}$:
  \beq
  e^{a(\eta)}_{0i}   &=& -\frac{1}{\eta}~ \e^{abc}_{} \left\{\phi F^i_{bc}(A)
  -\left( \phi -\eta \th\right) D^{}_{[b}(A) K^i_{c]} - \left( \frac{(\eta^2_{} -1)\phi +2\eta \th }{\eta}\right)
  \e^{ijk}_{} K^j_b K^k_c\right\} \nonumber \\
  {\tilde e} ^{a(\eta)}_{0i}   &=& -\frac{1}{\eta}~ \e^{abc}_{} \left\{\th F^i_{bc}(A)
  -\left( \th  +\eta \phi\right) D^{}_{[b}(A) K^i_{c]} - \left( \frac{(\eta^2_{} -1)\th -2\eta \phi }{\eta}\right)
  \e^{ijk}_{} K^j_b K^k_c\right\} ~~~
  \eeq
  
  The  $\chi^a_i$ constraints  (\ref{chi}) are of
  particular interest. To study their effect, we
  note that $(A^i_a, ~ {\hat E}^b_j)$ and $(K^i_a, {\hat F}^b_j)$ are
  canonically conjugate pairs. They have accordingly the standard
  Poisson brackets. From these, using  the relation (\ref{Econd}) expressing  $E^a_i $  
  in terms of   ${\hat E}^a_i$,   $A^i_a$ and $K^i_a$, as indicated in the Appendix,
  the following Poisson brackets can be calculated with respect to 
  phase variables $(A^i_a, {\hat E}^a_i)$ and $( K^i_a, {\hat F}^a_i)$: 
  \beq
  &&\left[A^i_a(x),~ E^b_j(y)\right]= \left[A^i_a(x), ~{\hat E}^b_j(y)\right] = \delta^i_j \delta^b_a ~\delta^{(3)}(x,y),\nonumber \\
  &&\left[ K^i_a(x),~ E^b_j(y)\right]   = 0~, ~~~~~~~\left[ E^a_i(x), ~E^b_j(y) \right] ~=~ 0  \nonumber
  \eeq 
  These then imply the Poisson bracket relations:
  \beq
  && \left[ \chi^a_i(x), ~A^j_b(y) \right] ~=~0~, ~~~~~~~\left[ \chi^a_i(x), ~K_b^j(y)
  \right]~ =~- \delta^i_j \delta^a_b \delta^{(3)}_{}(x,y)~, \nonumber \\
  &&\left[ \chi^a_i(x), ~E^b_j(y) \right] ~=~0~, ~~~~~~~ \left[\chi^a_i(x), ~\chi^b_j(y) \right]
  ~=~0  
  \eeq
  
 Using these, we notice that the Poisson brackets
 of Hamiltonian constraint $H$ and   $\chi^a_i$ are  non-zero. Requiring
 ~$\left[ \chi^a_i(x), H(y) \right] \approx0$ leads us  to the secondary constraints as: 
 \beq 
 t^a_i - \left(\frac{1+\eta^2_{}}{\eta^2_{}} \right)\left\{\eta\e^{ijk}_{}
   D^{}_b(A) \left({\sqrt E} E^a_jE^b_k \right) +{\sqrt E}E^{[a}_j
   E^{b]}_i K^j_b \right\}~ \approx~ 0 \nonumber 
   \eeq 
   which can be rewritten as: 
 \beq 
 t^a_i - \left(\frac{1+\eta^2_{}}{\eta^2_{}}
 \right)\e^{abc}_{} \left\{ \eta D^{}_b(A)v^i_c - \e^{ijk}_{} K^j_b
   v^k_c \right\} ~\approx ~ 0 \nonumber 
   \eeq 
   Next, since from (\ref{tau2}) and (\ref{tau3}),   $t^a_i \approx
 \tau^a_i ~\equiv~ \e^{abc}_{} \left\{ \eta D^{}_b(A)v^i_c -
   \e^{ijk}_{} K^j_b v^k_c \right\}$, this implies $t^a_i ~\approx~0$.
 Thus we have the  constraints: 
 \beq 
 \e^{abc}_{} \left\{ \eta
   D^{}_b(A)v^i_c - \e^{ijk}_{} K^j_b v^k_c \right\} ~\approx~ 0 \nonumber 
 \eeq 
These can be solved for the extrinsic curvature
 $K^i_a$ and recast as the following secondary constraints: 
 \beq\label{kconstraint}
 \psi^i_a &\equiv & K^i_a - \kappa^i_a (A,E)~\approx~ 0~, \nonumber \\
 \kappa^i_a (A,E) &\equiv & \frac{\eta}{2} ~\e^{ijk}_{} E^j_a
 D^{}_b(A)E^b_k  \nonumber \\ &&~~~~~~~~~~~-\frac{\eta}{2E} E^k_a \e^{bcd}_{} \left\{
   E^k_bD^{}_c(A)E^i_d + E^i_b D^{}_c(A)E^k_d - \delta^{ik}_{}E^m_b
   D^{}_c(A) E^m_d \right\} ~~~~~~~~~~\label{psi}
   \eeq 
   These are additional constraints and have the important property that these 
 form second class pairs with the constraints $\chi^a_i$ of (\ref{chi}):
 \beq 
 \left[\chi^a_i(x), ~\psi^j_b(y)\right] ~=~-\delta^a_b \delta^j_i \delta^{(3)}(x,y) 
 \eeq 
 
 To implement these second class constraints,
 $\chi^a_i$ and $\psi^i_a$, we need to go over
 from Poisson brackets to the corresponding Dirac brackets and then
 impose the constraints strongly, $\psi_a^i =0$ (which also implies $t^a_i =0$)   and $\chi_i^a=0$,
 in accordance with  Dirac theory of constrained Hamiltonian systems.
 As outlined in the Appendix,   {\it the Dirac brackets of 
 fields $A^i_a$ and $E^a_i$ turn out to be the same as their Poisson brackets; these are displayed  in (\ref{Dirac1}).
 On the other hand, those for $(A^i_a,~ {\hat E}^a_i; ~K_a^i, ~{\hat F}^a_i)$ are 
   different; these have been  listed in 
 (\ref{Dirac2}) and (\ref{Dirac3}).}

Finally, after implementing   these second class constraints, we have the Lagrangian density in the time-gauge as:
\beq
{\cal L}=\hat{E}^{a}_{i}\del_{t}A_{a}^{i}~+~\hat{F}^{a}_{i}\del_{t}K_{a}^{i}~ - {\cal H}  \label{L8}
\eeq
with  the Hamiltonian density
\beq
{\cal H} &=& NH+ N^a_{}H^{}_a + \frac{1}{2}~ \e^{ijk}_{}\w^{jk}_t G^{rot}_k
 \eeq
 and a set of seven first class constraints: 
\beq
&& G^{rot}_i  ~\equiv~  \eta ~D^{}_a(A){\hat E}^a_i ~+~\e^{ijk}_{} K^j_a {\hat F}^a_k ~\approx~ 0 \nonumber \\
&& H^{}_a ~\equiv~ {\hat E}^b_i F^i_{ab}(A) ~+~ {\hat F}^b_i D^{}_{[a}(A)
K^i_{b]} ~- ~K^i_a D^{}_b(A) {\hat F}^b_i
- {\eta}^{-1}_{} ~G^{rot}_i K^i_a ~\approx~ 0 \nonumber \\
&& H ~\equiv~  \frac{\sqrt E}{2\eta} \e^{ijk}_{}   E^a_i  E^b_j   F^k_{ab}(A)  -\left(
  \frac{1+\eta^2_{}}{2\eta^2_{}}\right){\sqrt E} E^a_i 
     E^b_j   K^i_{[a} K^j_{b]} + \frac{1}{\eta}~\partial^{}_a\left({\sqrt E} G^{rot}_k E^a_k\right)
\approx ~0 ~~~~~~~~~\label{finalconstraints}
\eeq 
with $E^a_i$ in the last equation given by:
$
 E^a_i=~ E^a_i ({\hat E}, A, K)\equiv {\hat E}^a_i + \frac{2}{\eta}  ~  {\tilde e}^{a(\eta) }_{0i}(A,K)   
$.
The fields $(A^i_a,~ {\hat E}^a_i,~ K^i_a,~ {\hat F}^a_i)$ have non-trivial  Dirac brackets as listed in
(\ref{Dirac2}) and (\ref{Dirac3}).
  The   second class
constraints $\chi^a_i$ and $\psi^i_a$   are now set  strongly equal to  zero: 
\beq 
&&K^i_a ~= ~\kappa^i_a (A,E)
~\equiv~  \frac{\eta}{2} ~\e^{ijk}_{} ~E^j_a D^{}_b(A)E^b_k -\frac{\eta}{2E} ~E^k_a
~\e^{bcd}_{} \left\{ E^k_bD^{}_c(A)E^i_d + E^i_b D^{}_c(A)E^k_d ~~~~~~~~~~~~\right. \nonumber \\
&&\left. ~~~~~~~~~~~~~~~~~~~~~~~~~~~~~~~~~~~~~~~~~~~~~~~~~~~~~~~~~~~~~~~~~~~~~~~~~~~~~~~~~~- \delta^{ik}_{}E^m_b 
D^{}_c(A) E^m_d \right\}\nonumber \\
&&{\hat F}^a_i ~= ~2\left( \eta + \frac{1}{\eta} \right) {\tilde e}^a_{0i} (A, K) 
  \eeq
 In writing the Hamiltonian constraint $H$ in (\ref{finalconstraints}) from (\ref{GHHconstraints}), we have    used  
  the   identity:
 \beq
  {\sqrt E} ~\e^{ijk}_{} E^b_i E^c_j \left( D^{}_b(A) K^k_c - \frac{1}{\eta}~ \e^{kmn}_{} K^m_b K^n_c \right)
 = -~\partial^{}_a \left( {\sqrt E} E^a_i G^{rot}_i\right)
 \eeq
which holds due to the  time gauge  relation $  ~E  E^a_i G^{rot}_i = \e^{abc}_{} E^i_b K^i_c$  ~with the constraints
 $K^i_a = \kappa^i_a (A, E)$ imposed strongly.

To evaluate the effect of   generators  (\ref{finalconstraints}) on various fields, we
need to use the Dirac brackets instead of the Poisson brackets. For
example, for the $SU(2)$ gauge generators,  using the results listed in the Appendix, we obtain: 
\beq
\left[G^{rot}_i(x), ~{\hat E}^a_j(y)\right]_D^{} &=& \e^{ijk}_{}{\hat E}^a_k ~\delta^{(3)}_{}(x,y)~, \nonumber \\
\left[G^{rot}_i (x), ~A^j_a (y)\right]_D^{} & = & - \eta \left(
  \delta^{ij}_{} \partial_a^{} ~ +  {\eta}^{-1}_{}~\e^{ikj}_{} A^k_a
\right) ~\delta^{(3)}_{}(x,y) 
\eeq 
reflecting  the fact $G^{rot}_i$ are
generators of $SU(2)$ transformations:   $A^i_a$ transform as the $SU(2)$   connection 
and   fields ${\hat E}^a_i $ as adjoint representations.
Besides, the fields  ${\hat F}^a_i, ~ K^i_a$ and $E^a_i$ also behave as adjoint representations under $SU(2)$ rotations:
 \beq
\left[G^{rot}_i (x), ~{\hat F}^a_j (y)\right]_D^{} &=& \e^{ijk}_{} {\hat F}^a_k ~\delta^{(3)}_{}(x,y) \nonumber \\
\left[G^{rot}_i (x), ~K^j_a (y) \right]_D^{} &=& \e^{ijk}_{} K^k_a ~\delta^{(3)}_{}(x,y)\nonumber \\
\left[G^{rot}_i (x), ~E^a_j (y)\right]_D^{} &=& \e^{ijk}_{} E^a_k
~\delta^{(3)}_{}(x,y)
 \eeq
  
 Similar discussion is
valid for the spatial diffeomorphism generators $H_a^{}$. The Dirac brackets
of $H_a^{}$ with various fields yield the Lie derivatives of these
fields respectively, modulo $SU(2)$ gauge transformations.

   As  stated earlier and demonstrated in the Appendix,   
     Dirac brackets for the fields $(A^i_a,~ {\hat E}^a_i; ~K_a^i, ~{\hat F}^a_i)$ are  
     different from their Poisson brackets (see  (\ref{Dirac1'}), 
      (\ref{Dirac2}) and (\ref{Dirac3})). This is so because the 
   transition from Poisson brackets to Dirac brackets, except for some special cases,  in general, does not preserve 
   canonical structure of the algebra \cite{mukunda}. When the second class constraints are imposed strongly,
   the algebraic structure of the   Dirac brackets of phase
   variables $(A^i_a,~{\hat  E}^a_i)$ of the final theory   is different from
   those of the phase variables $(A^i_a,~   E^a_i)$ of the standard canonical  theory. Thus the  
   variables $(A^i_a,~{\hat  E}^a_i)$   are not related to    $(A^i_a,~ E^a_i)$   through
   a canonical transformation.   However, it is possible to construct   a set of new phase space field variables
   whose  Dirac bracket algebra has the same structure as that of the standard canonical variables  $(A^i_a,~ E^a_i)$.
   
   In fact, in general, for theories with second class constraints as is the case here, instead of the
   ordinary  canonical transformations, what is relevant are the Gitman  D-transformations, which 
   preserve the form invariance of Dirac brackets and  equations of motion \cite{gitman}. Thus,
   in the present context also, new phase variables can be constructed through these D-transformations.
   These transformations change both the gauge fields as well as their conjugate momentum fields.
   This procedure finally leads to the phase variables:
   \beq
        E'^a_i(x) ~=~ \sum^{\infty}_{n=0} ~\frac{1}{n!}~ D'^{(n)a}_i(x)~,
         ~~~~~~~~~A'^i_a(x) ~=~ \sum^{\infty}_{n=0} ~\frac{1}{n!}~ C'^{(n)i}_a(x) \label{DT1}
   \eeq
   where
   \beq
   D'^{(0)a}_i(x) ~\equiv~ E^a_i(x)~,~~~~~~~~~~~~~~~~~~~~
C'^{(0)i}_a(x) ~\equiv~ A^i_a(x) \label{DT2}~~~~~~~~~~~
   \eeq
   and other $D'^{(n)}$ and $C'^{(n)}$ are   recursively  constructed using Dirac brackets as follows:
\beq
   &&D'^{(n+1)a}_i(x) = \int d^3_{}z ~  {\hat F}^b_l(z) \left[ K^l_b(z), ~ D'^{(n)a}_i(x)\right]_D^{} - 
   ~\frac{2}{\eta}~ \int d^3_{}z ~{\tilde e}^{(\eta)a}_{0l}(z) \left[ A^l_b(z),~ D'^{(n)a}_i(x) \right]^{}_D   
   \nonumber \\
         &&C'^{(n+1)i}_a(x) = \int d^3_{}z ~{\hat F}^b_l(z) \left[ K^l_b(z),~ C'^{(n)i}_a(x)\right]_D^{}~  
   - 
   ~\frac{2}{\eta}~ \int d^3_{}z ~{\tilde e}^{(\eta)a}_{0l}(z) \left[ A^l_b(z),~ C'^{(n)i}_a(x)\right]_D^{} \nonumber \\
   &&~~~~~~~~~~~~~~~~~~~~~~~~~~~~~~~~~~~~~~~~~~~~~~~~~~~~~~~~~~~~~~~~~~~~~~~~n=0,~ 1,~2,~3,.....  ~\label{DT3}
   \eeq
   In particular,
   \beq
   D'^{(1)a}_i(x) = {\hat F}^a_i(x)~-~\frac{2}{\eta} ~{\tilde e}^{(\eta)a}_{0i}(A,K;x)  , ~~~
   C'^{(1)i}_a(x)=  - \int d^3_{}z~ {\hat F}^b_l(z) 
  ~ \frac{\delta \kappa^l_b(A,E;z)}{\delta E^a_i(x)}~~
   ~\label{DT4}
   \eeq
   
    The  new variables $(A'^i_a,~ E'^a_i)$ are functions of the phase variables $(A^i_a, ~{\hat E}^a_i; ~K^i_a,~ {\hat F}^a_i)$ of the theory
 described above and can be checked to satisfy the Dirac bracket relations:
    \beq
    \left[ A'^i_a(x),~ E'^b_j(y)\right]^{}_D = \delta^j_i \delta^b_a \delta^{(3)}_{}(x,y),
    ~~\left[ A'^i_a(x),~ A'^j_b(y)\right]^{}_D = 0, ~~\left[E'^a_i(x),~ E'^b_j(y)\right]^{}_D =0~~~\label{DT5}
    \eeq
 As is expected under D-transformations,  these relations  reflect  the fact that the 
 algebraic structure  of    Dirac brackets for  the fields  $(A^i_a,~ E^a_i)$  as represented by (\ref{Dirac1}) has been preserved.
   After the second class 
   constraints, $\chi^a_i$ and $ \psi^i_a$,  are implemented,   $(A'^i_a,~ E'^a_i)$ 
          are
    related to the phase variables $(A^i_a,~ E^a_i)$  
   through an ordinary canonical transformation.
   
   We have not presented  many details of the construction of these new phase variables 
   above. Instead,
     in the next section,  we shall present, through an equivalent procedure,  an elaborate  construction of the new phase 
   variables  in the theory where second class constraints are already imposed strongly.
   This will be done by   a direct canonical transformation of the phase variables $(A^i_a,~ E^a_i)$ of the standard 
   canonical theory. The new canonical  variables so obtained will be shown to be equal to the  
   fields $A'^i_a$ and $E'^a_i$ above, when the second class constraints $\chi^a_i$ and $\psi^i_a$ are imposed.

\section{Canonical transformations and new phase   variables}
Adding the Nieh-Yan term to   Hilbert-Palatini Lagrangian density,  in the time gauge, leads to  a change of
  phase variables \cite{date}, from the ADM variables    $(\kappa^i_a, ~ E^a_i)$ to new variables $( A^i_a, ~ E^a_i)$. This change is just a 
 canonical transformation. Further inclusion of the Pontraygin and Euler densities   results in a  theory
 which can also be described in terms of canonically transformed phase variables. In the following,
 we shall develop such a description explicitly.  

We start with the standard  canonical theory constructed from the Lagrangian
density containing the Hilbert-Palatini term and the Nieh-Yan density as in (\ref{L1}) with
$\th=0$ and $\phi=0$. This is described, after partial gauge fixing (the time gauge), where the second class constraints 
are imposed, in terms the $SU(2)$ gauge fields $A^i_a$ and their  conjugates, densitized triads  $E^a_i$,  by
 the Lagrangian density
 \beq
&& {\cal L  }_1^{} ~=~ E^a_i\partial^{}_t A^i_a - {\cal H}  \nonumber \\
  && {\cal H} ~=~ \frac{1}{2}~\e^{ijk}_{}\w^{ij}_t G^{rot}_k ~+~ N^a_{} H_a ~+~ N H \nonumber \\
 && G^{rot}_i (A,E)~=~ \eta D^{}_a(A)  E^a_i~, ~~~~~~~~ H^{}_a (A,E)~=~ E^b_i F^i_{ab}(A) ~-~ \eta^{-1}_{} \kappa^i_a G^{rot}_i
 \nonumber \\
 &&H(A,E)~=~ \frac{\sqrt E}{2\eta} ~\e^{ijk}_{} E^a_i E^b_j \left( F^k_{ab}(A) - \frac{1+\eta^2_{}}{\eta} 
 ~\e^{kmn}_{} \kappa^m_a \kappa^n_b \right) ~+~ \frac{1}{\eta} ~\partial_a^{}\left({\sqrt E} G^{rot}_k E^a_k\right)~~~~
 \label{ABI}~~~~
 \eeq
where the extrinsic curvature $\kappa^i_a(A,E)$ is given in terms of $A^i_a$ and $E^a_i$ through (\ref{A3}).
Canonical pairs of the phase   variables $(A^i_a,~ E^a_i)$ obey the standard Poisson bracket relations:
\beq
\left[A^i_a(x),~ E^b_j(y)\right] = \delta^i_j~ \delta^b_a~\delta^{(3)}_{}(x,y)~, ~~~~ \left[A^i_a(x),~ A^j_b(y)\right] = 0 ~, ~~~~\left[E^a_i(x),~ E^b_j(y)\right] = 0~~~~~~\label{SPB1}
\eeq

  Next, we   add   the Pontraygin and Euler densities (\ref{Po2}, \ref{Eu2}), 
which are total divergences, $\frac{\th}{4} I_P^{} + \frac{\phi}{4}I^{}_E$
$= \partial^{}_\m J^\m_{}$, to ${\cal L}_1^{}$ above. The resulting  Lagrangian density, ignoring the spatial derivative part,  is
\beq
{\cal L }_2^{} ~=~ E^a_i\partial^{}_t A^i_a + \partial^{}_t J^t_{} - {\cal H} \label{Lag2}
\eeq
Inclusion of the  time derivative term  here is equivalent to a canonical transformation on  the phase space which can be
 constructed using  $J^t$. For this purpose, we first express $J^t_{}$ as 
a function of the phase variables $A^i_a$ and $E^a_i$:
\beq
  J^t_{} (A, \kappa(A,E))&= &\frac{\th}{\eta^2_{}}~   \e^{abc} \left\{ A^i_a F^i_{bc}  - \frac{1}{3\eta}~ \e^{ijk}_{} A^i_a A^j_b A^k_b\right\}
\nonumber \\
&&-~\frac{1}{\eta^2_{}}\left(\th+\eta\phi\right)  ~\e^{abc}_{} \left\{ \kappa^i_a F^i_{bc}  + A^i_a \left( D^{}_{[b}(A)
\kappa^i_{c]} - \frac{1}{\eta} ~\e^{ijk}A^j_b \kappa^k_c \right) \right\} \nonumber \\
 &&+  \frac{1}{\eta^2_{}}\left\{(1-\eta^2_{}) \th + 2\eta \phi\right\}\e^{abc}_{} \kappa^i_a D^{}_{[b}(A) \kappa^i_{b]} \nonumber \\
&&+\frac{2}{3\eta^3{}} \left\{ (3\eta^2_{} -1) \th - \eta( 3-\eta^2_{}) \phi \right\} \e^{abc}_{} \e^{ijk}_{}
\kappa^i_a \kappa^j_b \kappa^k_c \label{J1}
\eeq
Generating functional for the canonical transformation is:
\beq
{\cal J}(A,E) ~ = ~\int d^3_{} z ~J^t(A(z), \kappa(A(z),E(z))) \label{J2}
\eeq
which has functional dependence on both gauge fields $A^i_a$ and their conjugates $E^a_i$. 
Following the standard procedure, ${\cal J}$ generates  the canonical transformations,
$(A^i_a(x),~ E^a_i(x))$ $\rightarrow $ $({\cal A}^i_a(x),~ {\cal E}^a_i(x))$,  where
the new phase variables are given in terms of Poisson bracket series as follows: 
\beq
{\cal A}^i_a(x) &=& A^i_a(x) + \left[ {\cal J},A^i_a(x)\right] + \frac{1}{2!} \left[{\cal J}, 
\left[ {\cal J}, A^i_a(x)\right] \right] +\frac{1}{3!}~ \left[ {\cal J}, \left[ {\cal J}, \left[ {\cal J},  
A^i_a(x)\right] \right] \right] + .....~~~~~~~~~~\nonumber \\
&\equiv& ~ e^{\cal J}~ A^i_a(x) ~e^{-{\cal J}} \nonumber \\
 {\cal E}^a_i(x)&= &E^a_i(x)+ \left[ {\cal J},E^a_i(x)\right] + \frac{1}{2!} \left[{\cal J}, 
\left[ {\cal J}, E^a_i(x)\right] \right] +\frac{1}{3!}~ \left[ {\cal J}, \left[ {\cal J}, \left[ {\cal J},  
E^a_i(x)\right] \right] \right] + .....\nonumber\\
&\equiv& ~ e^{\cal J}~ E^a_i(x)  ~e^{-{\cal J}}  \label{newvar1}
\eeq
Alternately, these relations may  be represented as:
\beq
~~{\cal A}^i_a(x) = \sum_{n=0} \frac{1}{n!}~ C^{(n)i}_a(x)~, ~~~~~~~
{\cal E}^a_i(x)= \sum_{n=0} \frac{1}{n!}~ D^{(n)a}_i(x)   ~~~~~~~~~~~~~~~~~~~~~~~~~~~~~ \label{newvar2}
\eeq
with
\beq
&&C^{(0)i}_a = A^i_a(x), ~~~~~~~~~~~~~~~~~~~~~ D^{(0)a}_i(x) = E^a_i(x), \nonumber \\
&& C^{(n)i}_a(x) =\left[{\cal J}, ~C^{(n-1)i}_a(x)\right], ~~~ D^{(n)a}_i(x) = \left[{\cal J}, 
~D^{(n-1)a}_i(x) \right],~~~ n=1,~2,~3, ..... ~~\label{newvar3}
\eeq 
The various terms can be evaluated recursively through the following formulae:
\beq
C^{(n)i}_a(x)&=& \int d^3_{}z~ \left( D^{(1)b}_l(z) \left[ A^l_b(z),~ C^{(n-1)i}_a(x)\right]
- C^{(1)l}_b(z)\left[ E^b_l(z),~ C^{(n-1)i}_a(x)\right] \right) ~~~~~~~~~~~\nonumber \\
D^{(n)a}_i(x)&=& \int d^3_{}z~ \left( D^{(1)b}_l(z) \left[ A^l_b(z),~ D^{(n-1)a}_i(x)\right]
- C^{(1)l}_b(z)\left[E^b_l(z),~ D^{(n-1)a}_i(x)\right] \right) ~~~\nonumber \\
&& ~~~~~~~~~~~~~~~~~~~~~~~~~~~~~~~~~~~~~n=1,2,3........
\eeq
where, using ${\cal J}$ from (\ref{J1}, \ref{J2}),
 \beq
C^{(1)i}_a(x) &= &\left[{\cal J}, ~A^i_a(x)\right] = -~\frac{2(1+\eta^2_{})}{\eta} \int d^3_{}u~ {\tilde e}^b_{0l}(\kappa;u) 
~\frac{\delta f^l_b(u)}{\delta E^a_i(x)}~, \nonumber \\
D^{(1)a}_i(x) &= & \left[{\cal J}, ~E^a_i(x)\right] = 2 e^{(\eta)a}_{0i}(\kappa;x)~~~~~~~~~~~\label{C1D1}
\eeq
Here $e^{(\eta)a}_{0i}(\kappa;x)  \equiv e^a_{0i}(\kappa;x) +\eta {\tilde e}^a_{0i}(\kappa;x)$ and the argument $\kappa$
is to indicate that these functions are given by (\ref{e(A,K)}) with  $K_a^i$  replaced by $\kappa^i_a(A,E)
=A^i_a +f^i_a(E)$ 
where   $f^i_a(E)$ are as in  (\ref{A3}). 

 By repeated use of Jacobi identity, it can be checked that the functions $C^{(n)}$ and $D^{(n)}$ of  (\ref{newvar3}) satisfy the Poisson bracket relations:
\beq
\sum^{n}_{l=0} \frac{1}{l!(n-l)!}~ \left[ C^{(l)i}_a(x),~ C^{(n-l)j}_b(y)\right]&=0 \nonumber \\
\sum^{n}_{l=0} \frac{1}{l!(n-l)!} ~  \left[ D^{(l)a}_i(x),~ D^{(n-l)b}_j(y)\right]&=0 \nonumber \\
\nonumber \\
\sum^{n}_{l=0} \frac{1}{l!(n-l)!} ~\left[ C^{(l)i}_a(x),~ D^{(n-l)b}_j(y)\right]&=&0~, ~~~n=1,2,3, ......~~~~~~~
\label{idCD}
\eeq
 
The Poisson bracket relations (\ref{SPB1})  imply,  by construction,
  same Poisson brackets  for  the new variables (\ref{newvar1}):
\beq
\left[{\cal A}^i_a(x),~ {\cal E}^b_j(y) \right] = \delta^b_a~ \delta^i_j ~\delta^{(3)}_{}(x,y)~, 
~~~~~\left[{\cal A}^i_a(x),~ {\cal A}^j_b(y)\right] =0~, ~~~~ \left[{\cal E}^a_i(x), ~ 
{\cal E}^b_j (y) \right] =0~~~~~~~
\eeq
where the Poisson brackets are evaluated with respect to the phase variables $(A^i_a,~ E^a_i)$.
This can be readily checked by using the identities (\ref{idCD}).

For a general analytic function $P(A,E)$ of the phase variables $A^i_a$ and $E^a_i$, the following relation holds:
\beq
P({\cal A}, {\cal E}) = e^{\cal J} P(A,E) e^{-{\cal J}}  
&\equiv& P(A,E) + \left[ {\cal J}, P(A,E)\right]
+ \frac{1}{2!} \left[{\cal J}, 
\left[ {\cal J}, P(A,E)\right] \right] \nonumber \\
&&~~~~~~~~~~~~~~~~~~~~~~~~~~~+\frac{1}{3!}\left[ {\cal J}, \left[ {\cal J}, \left[ {\cal J},  
P(A,E)\right] \right] \right] + .....~~~~~~~~\label{P(AE)}
\eeq
Further    ${\cal J}$ of (\ref{J1} , \ref{J2}) written as a functional of   $ (A^i_a,~ E^a_i)$  
and $ ({\cal A}^i_a,~ {\cal E}^a_i)$ is form invariant:
\beq
{\cal J}({\cal A},{\cal E}) ~=~ {\cal J}(A,E) \label{J3}
\eeq

The converse relations expressing   $A^i_a$ and $E^a_i$ in terms of the transformed variables 
are:
\beq
 A^i_a(x) &=& {\cal A}^i_a(x) -\left[ {\cal J},{\cal A}^i_a(x)\right] + \frac{1}{2!} \left[{\cal J}, 
\left[ {\cal J}, {\cal A}^i_a(x)\right] \right] - \frac{1}{3!}\left[ {\cal J}, \left[ {\cal J}, \left[ {\cal J},  
{\cal A}^i_a(x)\right] \right] \right] + .....~~~~~~~~~~\nonumber \\
&\equiv& ~ e^{-{\cal J}}~ {\cal A}^i_a(x) ~e^{{\cal J}} \nonumber \\
  E^a_i(x)&= &{\cal E}^a_i(x) - \left[ {\cal J},{\cal E}^a_i(x)\right] + \frac{1}{2!} \left[{\cal J}, 
\left[ {\cal J}, {\cal E}^a_i(x)\right] \right] -\frac{1}{3!} \left[ {\cal J}, \left[ {\cal J}, \left[ {\cal J},  
{\cal E}^a_i(x)\right] \right] \right] + .....\nonumber\\
&\equiv& ~ e^{-{\cal J}}~ {\cal E}^a_i(x)  ~e^{{\cal J}}  \label{oldvar1}
\eeq
where ${\cal J}$ is written as a functional of ${\cal A}^i_a$ and ${\cal E}^a_i$  (refer (\ref{J3}))
and Poisson brackets are evaluated
with respect to these new variables.

Next, we evaluate the following:
\beq
~~~~~~\int d^3_{}x~{\cal E}_i^a(x) \partial^{}_t {\cal A}_a^i(x) = \sum^{\infty}_{n=0} F^{(n)}_{} , ~~~~~~~ F^{(n)} \equiv \sum_{l=0}^n \frac{1}{l!(n-l)!} \int d^3_{}x~ D^{(l)a}_i(x) \partial^{}_t C^{(n-l)i}_a(x)\nonumber
\eeq
It is straight forward to check:
\beq
F^{(0)}& = &\int d^3_{}x~ E^a_i(x) \partial^{}_t A^i_a(x) , ~~~~~~~~~~~~
F^{(1)} = \partial^{}_t G^{(0)} + \partial^{}_t {\cal J} \nonumber \\
F^{(n)} &=& \frac{1}{n!}~ \partial^{}_t G^{(n-1)}, ~~~~~~~~~~~ n =2,~3,~4, .......
\eeq
where 
\beq
G^{(n)} &\equiv& \left[ {\cal J}, ~ G^{(n-1)}\right] = \sum^n_{l=0} \frac{n!}{l!(n-l)!} 
\int d^3_{}x~ D^{(l)a}_i(x)C^{(n+1-l)i}_a(x)~,  ~~~~   n=1,2,3.....~~~~~~~~~~~ \nonumber\\
G^{(0)} &\equiv& \int d^3_{}x~ E^a_i(x) C^{(1)i}_a(x) \label{G}
\eeq
To obtain this result, the following helpful identities may be used:
\beq
 \sum^{n-1}_{l=0} \frac{n!}{l!(n-1-l)!}\int d^3_{}x \left( D^{(n-l)a}_i  \delta C^{(l)i}_a  
- C^{(n-l)i}_a  \delta D^{(l)a}_i \right)=0,~~~~ n=2,3,4.....~~~~~
\eeq
which can be derived recursively by taking Poisson brackets with ${\cal J}$. 

Further, using expression 
for $C^{(1)i}_a$  from (\ref{C1D1}) and eqns.  (\ref{deltaf}, \ref{SA}) of the Appendix, 
the following relation  can be obtained:
\beq
E^a_i(x)C^{(1)i}_a(x) = (1+\eta^2_{}) ~\e^{ijk}_{}~\partial_a^{} \left( {\tilde e}^b_{0i}(\kappa;x)E^j_b(x)E^a_k(x)\right)
\nonumber
\eeq
which in turn implies $G^{(0)} \equiv \int d^3_{}x~ E^a_i(x) C^{(1)i}_a (x) =0$ and hence all the $G^{(n)}$ of (\ref{G}) 
are zero, thus  leading to the result:
\beq
\int d^3_{}x~{\cal E}_i^a(x) \partial^{}_t {\cal A}_a^i(x) ~= ~\int d^3_{}x~ E^a_i(x) \partial^{}_t A^i_a(x)
~~+ ~~\partial^{}_t {\cal J} \label{EA}
\eeq

Since the generating functional ${\cal J}$, as given by (\ref{J1}, \ref{J2}), is invariant under 
small $SU(2)$ gauge transformations and spatial diffeomorphisms generated 
respectively by $G^{rot}_i(A,E)$ and $H_a^{}(A,E)$ of (\ref{ABI}):
\beq
\left[ {\cal J}, ~G^{rot}_i(A,E)\right] =0~, ~~~~~~~~~~~ \left[{\cal J}, ~H_a(A,E)\right] =0 \nonumber 
\eeq
Consequently,   $G^{rot}_i$ and  $H^{}_a$ written in terms of the phase variable $(A^i_a, ~E^a_i)$ and 
 $({\cal A}^i_a, {\cal E}^a_i)  $ are form invariant:
 \beq
 G^{rot}_i({\cal A}, {\cal E}) &=& e^{\cal J}~G^{rot}_i(A,E) ~e^{-{\cal J}} = ~G^{rot}_i(A,E)~ \nonumber \\
 H^{}_a({\cal A}, {\cal E}) &= &e^{\cal J}~H^{}_a(A,E) ~e^{-{\cal J}} = ~H^{}_a(A,E)
 \eeq
On the other hand, for the Hamiltonian constraint we have:
\beq 
H(A,E) &=& e^{-{\cal J}({\cal A}, {\cal E})}~H({\cal A}, {\cal E}) ~e^{{\cal J}({\cal A}, {\cal E})}  \nonumber \\
&=& H({\cal A}, {\cal E}) ~-~\left[{\cal J}, H({\cal A}, {\cal E})\right] ~+~\frac{1}{2!} \left[{\cal J}, \left[ {\cal J}, 
H({\cal A}, {\cal E})\right]\right]  \nonumber \\
&&~~~~~~~~~~~~~~~~~~~~~~~~~~~~~~~~
 -~\frac{1}{3!}\left[{\cal J}, \left[{\cal J}, \left[ {\cal J}, 
H({\cal A}, {\cal E})\right]\right]\right]+ .........  
\eeq
where the Poisson brackets are   with respect to phase variables $({\cal A}^i_a, ~{\cal E}^a_i)$.

This detail discussion,   finally allows us to write the theory based on the Lagrangian density (\ref{Lag2})
in terms of the new phase variables as:
\beq
{\cal L}_2   &= &~{\cal E}^a_i\partial_t^{} {\cal A}^i_a ~-~ {\hat{\cal H}} \nonumber \\
{\hat {\cal H}}&=& \frac{1}{2}\e^{ijk}_{}\w^{ij}_t G^{rot}_k({\cal A}, {\cal E}) + N^a_{} H^{}_a({\cal A}, {\cal E})
+ N{\hat H}({\cal A}, {\cal E})
\eeq
where
\beq
G^{rot}_i({\cal A}, {\cal E})& = &\eta D^{}_a ({\cal A}) {\cal E}^a_i \nonumber \\
H^{}_a({\cal A}, {\cal E})&=& {\cal E}^b_i {\cal F}^i_{ab}({\cal A}) - \eta^{-1}_{} G^{rot}_i({\cal A}, {\cal E})
\kappa_a^i({\cal A}, {\cal E}) \nonumber \\
{\hat H}({\cal A}, {\cal E})&=&e^{-{\cal J}({\cal A}, {\cal E})}~H({\cal A}, {\cal E}) ~e^{{\cal J}({\cal A}, {\cal E})} 
\eeq

The new variables $({\cal A}^i_a, ~{\cal E}^a_i)$ obtained here are related to the variables
$(A'^i_a, ~E'^a_i)$  of eqns. (\ref{DT1}-\ref{DT5})  of Sec.IV derived by the Gitman  D-transformations.
When the second class constraints $\chi^a_i$ and $\psi^i_a$ there are  implemented,
  $(A'^i_a, ~E'^a_i)$ collapse to $({\cal A}^i_a, ~{\cal E}^a_i)$:
\beq
A'^i_a (\chi =0, \psi  =0) = {\cal A}^i_a~, ~~~~ E'^a_i (\chi =0, \psi  =0) = {\cal E}^a_i
\eeq
This is so because each of the terms in (\ref{DT1}) and (\ref{newvar2}) coincide:
\beq
C'^{(n)i}_a(\chi=0, \psi=0) = C^{(n)i}_a~, ~~~~~~ D'^{(n)a}_i(\chi=0, \psi=0) = D^{(n)a}_i
\eeq

This completes our discussion of the canonical transformation  
to new variables $({\cal A}^i_a,~ {\cal E}^a_i)$ obtained by adding the Pontraygin and Euler densities to the standard  canonical
  theory of gravity described in terms of the phase variables $(A^i_a, ~ E^a_i)$. 

\section{Summary and Concluding remarks}
We have developed the canonical Hamiltonian formulation of gravity
theory with all the three topological terms of the Lagrangian density
(\ref{L1}) as an $SU(2)$ gauge theory with Barbero-Immirzi parameter
$\g = \eta^{-1}_{}$ as its coupling constant.  In time-gauge, the
theory containing only the Nieh-Yan topological term ($\th =0$,
$\phi=0$) developed earlier in ref.\cite{date}, is described by real
$SU(2)$ gauge fields, $A^i_a = \w^{0i}_a + \eta {\tilde {\w}}^{0i}_a$,
and densitized triads $E^a_i$ as their conjugate momentum fields.
This coincides with the standard $SU(2)$ gauge theoretical canonical
formulation of the theory of gravity \cite{rov}.  When the Pontryagin
and Euler terms are also included, there is a formulation of the
theory which retains the gauge fields $A^i_a$ (independent of the
topological parameters $\th$ and $\phi$) as the canonical fields, but,
their conjugate momentum fields are modified from $E^a_i$ to ${\hat
  E}^a_i \equiv E^a_i - 2 {\eta}^{-1}_{}~ {\tilde e}^a_{0i}(A,K) +2
e^a_{0i}(A,K) $ developing dependence on $\th$ and $\phi$. Further,
for the case with $\th=0$ and $\phi=0$, the momentum conjugate to
extrinsic curvature $K^i_a$ is zero. Here, in the most general case,
it is non-zero, represented by ${\hat F}^a_i$ which depends on other
fields through the $\chi^a_i$ constraints (\ref{chi}).  In addition,
it also depends on the topological parameters $\th$ and $\phi$.
Associated with $\chi^a_i$, we have a set of secondary constraints
$\psi^i_a$ of (\ref{psi}) which expresses the fact that extrinsic
curvature $K^i_a$ is not an independent field.  These constraints,
$(\chi^a_i, ~\psi^i_a)$, form second class pairs which are implemented
by going over to the Dirac brackets from Poisson brackets. The theory
is described by seven first class constraints: the $SU(2)$ gauge
constraints $G^{rot}_i$, spatial diffeomorphism constraints $H^{}_a$
and Hamiltonian constraint $H$ as listed in (\ref{finalconstraints}).
In this formulation, however, the Dirac brackets for the phase
variables $(A^i_a(x),~ {\hat E}^a_i(x))$ do not possess the same
algebraic structure as those for the canonical variable $(A^i_a(x), ~
E^a_i(x))$ of the standard theory.  Even after the second class
constraints, $\chi^a_i=0,$ $ ~ \psi^i_a=0$, are imposed, there is no
canonical transformation that relates the set $(A^i_a,~ {\hat E}^a_i)$
to $(A^i_a,~ E^a_i)$.  However, it is possible to construct another
Hamiltonian formulation in terms of new canonical variables $({\cal
  A}^i_a,~ {\cal E}^a_i)$ which indeed are related to the standard
variables $(A^i_a,~ E^a_i)$ through a canonical transformation. Here
both the gauge fields as well as their conjugate momentum fields, as
represented in (\ref{newvar1}), are changed and these depend on all
the topological parameters, $\eta$, $\th$ and $\phi$.

From this classical Hamiltonian formulation described in terms of
$({\cal A}^i_a,~ {\cal E}^a_i)$, we can go over to the quantum theory
by replacing the Poisson brackets by commutators of corresponding
operators in the usual fashion. We already have some evidence that
Barbero-Immirzi parameter $\eta^{-1}_{}$ is relevant in the quantum
theory.  For example, it appears in the spectrum of area and volume
operators \cite{ajrs} and also in the black hole entropy
\cite{entropy}.  How other parameters, $\th$ and $\phi$, will be
reflected in the quantum theory is an open question requiring deeper
study.

The analysis presented in the present article is for pure gravity
without matter couplings. Inclusion of matter, such as fermions, spin
$1/2$ or spin $3/2$ (supergravity), may be achieved through standard
minimal couplings.  All the topological densities in the Lagrangian
are described in terms of geometric quantities only. Their presence
does not change the classical equations of motion even with matter.  A
Hamiltonian formulation, in the time gauge, can again be set up in
terms of a real $SU(2)$ gauge theory with $\eta^{-1}_{}$ as its
coupling constant.

\acknowledgments
Discussions with Ghanashyam Date are gratefully acknowledged. R.K.K also 
acknowledges the support of   Department of Science and  Technology, Government of India through a J.C. Bose Fellowship.

\appendix*
\section{Poisson and Dirac brackets} In the time-gauge Lagrangian density (\ref{L8}), the fields 
$(A^i_a,~ {\hat E}^a_i)$ and $(K^i_a, ~ {\hat F}^a_i)$ are canonical pairs which have the standard Poisson bracket relations:
\beq
[A^i_a(t, {\vec x}),~ {\hat E}^b_j(t, {\vec y})] ~= ~\delta^i_j \delta^b_a  ~\delta^{(3)}_{}({\vec x}, {\vec y})~, ~~~~~
[K^i_a(t, {\vec x} ),~ {\hat F}^b_j( t, {\vec y})] ~=~ \delta^i_j \delta^b_a  ~\delta^{(3)}_{}( {\vec x}, {\vec y}) ~~~~
\label{A1}
\eeq
and all other brackets amongst these fields are zero. Thus the Poisson bracket for any two arbitrary fields $P$ and $Q$     is given by:
\beq
\left[ P(x),~ Q(y)\right] &=& \int d^3_{}z \left( \frac{\delta P(x)}{\delta A^i_a(z)} ~\frac{\delta Q(y)}{\delta{\hat E}^a_i(z)}
~-~ \frac{\delta P(x)}{\delta {\hat E}_i^a(z)}~ \frac{\delta Q(y)}{\delta A^i_a(z)} \right) \nonumber \\
&&~~~~+~\int d^3_{}z  \left( \frac{\delta P(x)}{\delta K^i_a(z)} ~\frac{\delta Q(y)}{\delta {\hat F}^a_i(z)}
~-~ \frac{\delta P(x)}{\delta {\hat F}_i^a(z)}~ \frac{\delta Q(y)}{\delta K^i_a(z)} \right)~~~~~~~~~~
\eeq

>From these, using
$E^a_i = E^a_i({\hat E}, A, K) \equiv {\hat E}^a_i +2 {\eta}^{-1}~{\tilde e}^{a(\eta) }_{0i}(A, K) $,
 we have the Poisson bracket relations
 \begin{eqnarray}
 \left[A^i_a(x),~ E^b_j(y)\right] ~= ~\left[A^i_a(x), ~{\hat E}^b_j(y)\right] ~= ~\delta^i_j \delta^b_a  ~\delta^{(3)}_{}(x, y), ~~
   ~~\left[K_{a}^{i}(x),~E^{b}_{j}(y)\right]~=~0  ~~~~~~
 \label{A2}
\end{eqnarray}
Using the expressions for $e^a_{0i}(A,K)$ and ${\tilde e}^a_{0i}(A,K)$ as functions of $A^i_a$ and $K^i_a$ as in (\ref{e(A,K)}), 
the following relations obtain:
\begin{eqnarray} 
&&\left[{\hat E}^a_i(x), E^b_j(y)\right]= \frac{2}{\eta}\left[{\hat E}^a_i(x), {\tilde e}^{b (\eta) }_{0j}(y)\right] 
 ~= -~\frac{4}{\eta^2_{}}~ 
\e^{abc}_{} \left\{  \theta D^{ij}_c   
  -\left(\frac{ \theta +\eta \phi}{\eta}\right) \e^{ikj}_{} K^k_c \right\} \delta^{(3)}_{} (x, y) \nonumber \\
&&\left[{\hat F}^a_i(x), E^b_j(y)\right]   =  \frac{2}{\eta} \left[{\hat F}^a_i (x), {\tilde e}^{b(\eta)}_{0j}(y)\right] \nonumber  \\
&&~~~~~~~~~~~~~~~~~~~~= \frac{4}{\eta^2_{}}~ \e^{abc}_{} \left\{  ( \th+\eta \phi)D^{ij}_c   
 -\left(\frac{(1-\eta^2_{})\th +2\eta \phi}{\eta} \right) \e^{ikj}_{} K^k_c 
\right\} \delta^{(3)}_{} (x, y) \nonumber \\
&&\left[{\tilde e}^{a (\eta) }_{0i}(x), E^b_j(y)\right]=  \left[{\tilde e}^{a(\eta)}_i(x), {\hat E}^b_j(y)\right] 
 ~= \frac{2}{\eta}~ 
\e^{abc}_{} \left\{  \theta D^{ij}_c   
  -\left( \frac{\theta +\eta \phi}{\eta}\right) \e^{ikj}_{} K^k_c \right\} \delta^{(3)}_{} (x, y) \nonumber \\
&&(1+\eta^2_{})\left[{\tilde e}^a_{0i}(x), E^b_j(y)\right]   =  (1+\eta^2_{}) \left[{\tilde e}^a_{0i} (x), {\hat E}^b_j(y)\right] \nonumber  \\
&&~~~~~~~~~~~~~~~~~~~~~~= \frac{2}{\eta }~ \e^{abc}_{}\left\{ ( \th+\eta \phi)D^{ij}_c   
 -\left(\frac{(1-\eta^2_{})\th +2\eta \phi }{\eta}\right) \e^{ikj}_{} K^k_c 
\right\} \delta^{(3)}_{} (x, y) ~~~~~~~~~~~~
 \end{eqnarray}
 \noindent where the $SU(2)$ gauge covariant derivative is: $
D^{ij}_c ~\equiv~ \delta^{ij}_{} \partial_c^{} + \eta^{-1}_{} \e^{ikj}_{} A^k_c \nonumber \label{A5}$.
These Poisson bracket relations imply for $E^a_i = E^a_i({\hat E}, A, K) \equiv {\hat E}^a_i +2 {\eta}^{-1}
~{\tilde e}^{a(\eta) }_{0i}(A, K) $:
\beq
\left[E^a_i(x),~ E^b_j(y)\right] ~=~0 \label{A6}
\eeq
 Now, using these Poisson bracket relations along with (\ref{A2}), yields:
\beq
 \left[\kappa_{a}^{i}(x),~E^{b}_{j}(y)\right]~=~\left[A_{a}^{i}(x),~E^{b}_{j}(y)\right]~=~\delta^{b}_{a} \delta^{i}_{j}
  ~\delta^{3}(x,y)
 \eeq
where   $\kappa^i_a (E,A)$ is given by (\ref{psi}) and can be rewritten explicitly as:
\beq
&&\kappa^i_a (A,E) ~=~ A^i_a ~+~ f^i_a (E)\nonumber \\
&&f_a^i (E) ~=~ \frac{\eta}{2} ~\e^{ijk}_{} E^j_a \partial^{}_b E^b_k -\frac{\eta}{2E}~ E^k_a ~\e^{bcd}_{} \left( E^i_b 
\partial^{}_c E^k_d + E^k_b \partial^{}_c E^i_d - \delta^{ik}_{} E^l_b \partial^{}_c E^l_d\right)  ~~~~~\nonumber \\
&& ~~~~~~~~=~ -\eta~ E^j_a ~\e^{bcd}_{} \left( v^i_b \partial^{}_c v^j_d 
~-~\frac{1}{2}~\delta^{ij}_{} ~v^r_b\partial^{}_c v^r_d \right) \label{A3}
\eeq
with   $v^i_a \equiv E^i_a / {\sqrt E}$. It is straight forward to check that  $f^i_a$ satisfy the identity:
\beq
 \e^{abc}_{} \left\{ \partial^{}_b E^i_c  -\partial_b^{}(\ln{\sqrt E})~ E^i_c -{\eta}^{-1}_{}\e^{ijk}_{} f^j_b E^k_c \right\} =0 \nonumber 
\eeq
  Equivalently, this relation can also  be written as:
 \beq
 \partial^{}_a E^a_i~-~ \eta^{-1}_{} \e^{ijk}_{} f^j_a E^a_k ~\equiv~ D^{}_a(A)E^a_i~-~\eta^{-1}_{}\e^{ijk}_{} \kappa^j_a E^a_k ~=~ 0
 \eeq
These relations can be used to calculate  the variation   $\delta f^i_a$  to be:
\beq
\delta f^i_a ~=~ {\cal S}^{il}_{ab} \delta E^b_l ~-~ \eta~ \partial^{}_c \left( {\cal A}^{~~~cil}_{ab} 
\delta E^b_l \right) + \frac{\eta}{2} \left( \partial^{}_c {\cal A}^{~~~cil}_{ab}\right) \delta E^b_l
\label{deltaf}
\eeq
with
\beq
{\cal S}^{il}_{ab}&=& -~\left( E^l_a f^i_b+E^i_bf^l_a\right) ~+~\frac{3}{4}  \left( E^i_af^l_b + E^l_bf^i_a\right) 
~-~\frac{1}{2}~E^m_aE^m_b\left(E^c_i f^l_c + E^c_l f^i_c \right) \nonumber \\
&~& +~\frac{1}{4}  \left( E^m_aE^l_b E^c_i + E^m_bE^i_aE^c_l\right) f^m_c ~+ ~\left( E^i_bE^l_a -E^i_aE^l_b
+ \delta^{li}_{} E^n_a E^n_b\right) E^c_m f^m_c \nonumber \\
&~& -~ \frac{\eta}{4}~ \e^{imk}_{} \left(\partial^{}_c E^m_a E^l_b -\partial^{}_c E^l_b E^m_a \right) E^c_k
~-~ \frac{\eta}{4}~ \e^{lmk}_{}\left( \partial^{}_c E^m_b E^i_a -\partial^{}_c E^i_a E^m_b \right) E^c_k
\nonumber \\
&~& +~ \frac{\eta}{2}~ \e^{ilk}_{} \left( \partial^{}_c E^m_a E^m_b - \partial^{}_c E^m_b E^m_a \right) E^c_k
\nonumber \\
{\cal A}^{~~~cil}_{ab} &=& \left( \e^{ilk}_{} E^m_a E^m_b - \frac{1}{2}~   \e^{imk}_{} E^m_a E^l_b 
+ \frac{1}{2} ~\e^{lmk}_{} E^m_b E^i_a \right) E^c_k \label{SA}
\eeq
Notice that ${\cal S}^{il}_{ab}$ and ${\cal A}^{~~~cil}_{ab}$ are respectively symmetric and antisymmetric 
under the interchange of the pair of indices $(a,i) $ and $ (b,l)$:
\beq
{\cal S}^{il}_{ab} ~=~ {\cal S}^{li}_{ba}~, ~~~~~~ {\cal A}^{~~~cil}_{ab}~=~-~ {\cal A}^{~~~cli}_{ba}
\eeq
These properties, immediately, lead to the relation:
\beq
\frac{\delta f^i_a(x)}{\delta E^b_l(y)} ~=~ \frac{\delta f^l_b(y)}{\delta E^a_i(x)}
\eeq

Next, using $\chi^a_i (x)\equiv {\hat F}^a_i(x) - \frac{2(1+\eta^2_{})}{\eta} ~{\tilde e}^a_{0i}(x)$
from (\ref{chi}),
 equations   (\ref{A1}) also imply the  following:
 \beq
 &&\left[\chi^a_i(x), {\hat E}^b_j(y)\right] = -~\frac{2(1+\eta^2_{})}{\eta}~\left[{\tilde e}^a_{0i}(x),  {\hat E}^b_j(y)
 \right] \nonumber \\
 &&~~~~~~~~=-~\frac{4}{\eta^2 _{}} \e^{abc}_{} \left\{ (\th +\eta \phi) D^{ij}_c - 
 \left(\frac{(1-\eta^2_{} )\th+2\eta\phi}{\eta}\right) \e^{ikj}_{} K^k_c \right\} \delta^{(3)}_{} (x, y) \nonumber \\
 &&\left[\chi^a_i(x), {\hat F}^b_j(y)\right] = -\frac{2(1+\eta^2_{})}{\eta}~\left[{\tilde e}^a_{0i}(x),  {\hat F}^b_j(y)
 \right] \nonumber \\
 &&~~~~~~~~=\frac{4}{\eta^2_{}} \e^{abc}_{} \left\{\left(  (1-\eta^2_{})\th +2\eta \phi \right) D^{ij}_c
 +\left( \frac{(3\eta^2_{}-1) \th -\eta (3-\eta^2_{}) \phi}{\eta} \right) \e^{ikj}_{} K^k_c \right\} \delta^{(3)}_{} (x, y)~~~~ \nonumber\\
   &&\left[\chi^a_i(x),~ A^j_b(y)\right]~=~ 0~, ~~~~~~~~~~ \left[\chi^a_i(x), ~ K^j_b(y)\right]~=~ -\delta^j_i \delta^a_b ~\delta^{(3)}_{} (x,y)
 \nonumber \\
 &&  (1+\eta^2_{}) \left[\chi^a_i(x), ~e^b_{0j}(y)\right] =  (1+\eta^2_{}) \left[{\hat F} ^a_i (x), ~ e^b_{0j}(y)\right] \nonumber \\
 && ~~~~~~~=~ \frac{2}{\eta}~ \e^{abc}_{} \left\{  \left( (1-\eta^2_{}) 
 \phi -2\eta \th\right) D^{ij}_c    
   ~+ \left(\frac{ \eta (3-\eta^2_{})\th +(3\eta^2_{} -1) \phi}{\eta} \right) 
 \e^{ikj}_{} K^k_c \right\}\delta^{(3)}_{} (x,y) \nonumber \\
 && (1+\eta^2_{}) \left[\chi^a_i(x), ~{\tilde e}^b_{0j}(y)\right] = (1+\eta^2_{}) \left[ {\hat F}^a_i(x),~ {\tilde e}^a_{0j}(y)\right] ~~\nonumber \\
 &&~~~~~~~~= \frac{2}{\eta}~ \e^{abc}_{} \left\{   \left( (1-\eta^2_{}) 
 \th +2\eta \phi\right) D^{ij}_c   ~+ \left( \frac{  (3 \eta^2_{}-1)\th -\eta(3-\eta^2_{} ) 
 \phi }{\eta}\right) 
 \e^{ikj}_{} K^k_c \right\}\delta^{(3)}_{} (x,y) ~~ \nonumber\\
 &&[\chi^a_i(x), ~{\tilde e}^{(\eta)b}_j(y)] = [{\hat F}^a_i(x), ~{\tilde e}^{(\eta)b}_j(y)]  \nonumber \\
 &&~~~~~~~~~= \frac{2}{\eta} ~\e^{abc}_{} \left\{  (\th +\eta \phi) D^{ij}_c - 
 \left(\frac{((1-\eta^2_{} )\th+2\eta\phi}{\eta} \right) \e^{ikj}_{} K^k_c \right\} \delta^{(3)}_{} (x, y)
 \eeq
 which further imply:
\beq
&& \left[\chi^a_i(x),~ E^b_j(y)\right]~=~ 0 ~, ~~~\left[\chi^a_i(x),~ \kappa^i_b(y)\right]~=~0, ~~~~ 
\left[\chi^a_i(x),~ \chi^b_j(y)\right]~=~0  ~~~
\eeq 

For $\psi_a^i \equiv K^i_a -\kappa^i_a(A, E)$ as given by (\ref{psi}),  using (\ref{A2}) and (\ref{A3}), we have  the following useful relations:
\begin{eqnarray}
&&\left[\psi^{i}_{a}(x), ~E^{b}_{j}(y)\right]~=~-\left[\kappa^i_a(x)~, E^b_j(y)\right]= -\delta^{b}_{a} \delta^{i}_{j} ~\delta^{3}(x,y), \nonumber\\
&&\left[\psi^i_a(x),~ A^j_b(y)\right]~=~ -\left[\kappa^i_a (x),~ A^j_b(y)\right] ~=~ \frac{\delta \kappa^i_a(x)}{\delta E^b_i(y)}  \nonumber \\
&&  
\left[\psi_{a}^{i}(x),~E_{b}^{j}(y)\right]~=~-~ \left[\kappa^i_a(x),~ E^j_b(y)\right] =~ ~E_{a}^{j}E_{b}^{i}~\delta^{3}(x,y), ~~~~~
~~~\nonumber \\
&& 
\left[\psi^{i}_{a}(x),~E(y)\right]~=~-~\left[\kappa^i_a(x),~ E(y)\right] ~=~E E_{a}^{i}~\delta^{3}(x,y)~~~~
\end{eqnarray}
 
The Poisson bracket relations among $\chi^a_i$ and $\psi^i_a$, obtained by using the properties listed  above,
 can be summarized as:
\beq
&&\left[\chi^a_i(x), \chi^b_j(y)\right]=0~,   ~~~\left[\chi^a_i(x), \psi^j_b(y)\right] =  
-\delta^a_b \delta^j_i~ \delta^{(3)}_{}(x,y)~ , ~~~\left[\psi^i_a(x), \psi^b_j(y)\right]= 
0  ~~~~~~~~~~~~~\label{A9}
\eeq
where  the last equation follows from the relation:
\beq
\left[ \kappa^i_a(x),~ \kappa^j_b(y)\right] ~=~ \left[ A^i_a(x),~f^j_b(y)\right] + \left[f^i_a(x), ~A^j_b(y) \right]  
 ~=~\frac{\delta f^j_b(y)}{\delta E^a_i(x)} ~-~  \frac{\delta f^i_a(x)}{\delta E^b_j(y)} ~=~ 0 ~~~~~~\label{A10}
\eeq
Here the Poisson brackets  involving $f^i_a(E)$ are calculated by
using their expressions as functions of $E^i_a$ as given by (\ref{A3}). The  identity (\ref{A10}) further implies 
the following Poisson bracket relations:
\beq
&&\left[ F^i_{ab}(x), ~ \kappa^j_d(y)\right] + \left[D^{}_{[a}(A)\kappa^i_{b]} (x),~ A^j_d(y) \right] =0 , \nonumber \\
&&\left[ F^i_{ab}(x), ~ D^{}_{[c}(A)\kappa^j_{d]}(y)\right] + \left[D^{}_{[a}(A)\kappa^i_{b]} (x),~ F^j_{cd}(y) 
\right] =0, \nonumber \\
&&\eta\left[D^{}_{[a}(A) \kappa^i_{b]}(x),  D^{}_{[c}(A)\kappa^j_{d]}(y) \right] + \left[ F^i_{ab}(x),
 \e^{jmn}_{} \kappa^m_c(y) \kappa^n_d(y) \right] 
+ \left[\e^{ikl}_{} \kappa^k_a(x) \kappa^l_b(x), F^j_{cd}(y) \right] =0 ~~~~ \nonumber \\
&&\left[ D^{}_{[a}(A)\kappa^i_{b]}(x), ~ \e^{jmn}_{} \kappa^m_c (y)\kappa^n_d(y)\right]
+ \left[ \e^{ikl}_{} \kappa^k_a (x)\kappa^l_b(x), ~ D^{}_{[c}(A)\kappa^j_{d]} (y)\right]=0,  
\eeq

To implement the second-class constraints $\chi^{a}_{i}\approx 0$ and
$\psi^{a}_{i}\approx 0$, we need to go over to the corresponding Dirac
brackets and then put $\chi^{a}_{i}= 0$ and $\psi^{a}_{i}= 0$
strongly. 
  From the Poisson bracket relations of these constraints (\ref{A9}),  the Dirac bracket of  
  any two fields C and D can be constructed to be:
\begin{eqnarray}
 \left[C,~D \right]_{D}^{} ~=~\left[C,~D \right] - \left[C,~\chi \right]~\left[\psi, ~D \right]+
 \left[C,~\psi \right]~\left[\chi, ~D \right]
\end{eqnarray}

Using the Poisson bracket relations listed above, it is straight forward to check that  the Dirac brackets amongst 
$A^{i}_{a}$ and $E^{a}_{i}$ are the same as their Poisson brackets:
\begin{eqnarray}
&&\left[E^{a}_{i}(x),E^{b}_{j}(y)\right]_D^{} =~\left[E^{a}_{i}(x),E^{b}_{j}(y)\right]~=~0~,  
 ~~~\left[A_{a}^{i}(x), A_{b}^{j}(y)\right]_D^{}  =~\left[A_{a}^{i}(x),A_{b}^{j}(y)\right]~=~0\nonumber  \\
&& \left[A_{a}^{i}(x),E^{b}_{j}(y) \right]_D^{} =~\left[A_{a}^{i}(x),E^{b}_{j}(y) \right]~=~\delta^{b}_{a} \delta^{i}_{j} 
~\delta^{3}(x,y)\label{Dirac1}
\end{eqnarray}
Also we note that,
\beq
&&\left[ K_{a}^{i} (x), E^{b}_{j} (y) \right]_D^{} = \left[\kappa_{a}^{i} (x), E^{b}_{j} (y) \right]_D^{}
  =\left[\kappa_{a}^{i}(x),E^{b}_{j}(y)\right]=[A_{a}^{i}(x),E^{b}_{j}(y)]
  =\delta^{b}_{a} \delta^{i}_{j} ~\delta^{3}(x,y) ,\nonumber \\
 &&\left[K_{a}^{i}(x),A_{b}^{j}(y)\right]_D^{}   = \left[\kappa_{a}^{i}(x),A_{b}^{j}(y)\right]_D^{}
  =\left[\kappa_{a}^{i}(x),A_{b}^{j}(y)\right]=\left[f_{a}^{i}(x),A_{b}^{j}(y)\right]
  = - ~\frac{\delta f^i_a(x)} {\delta E^b_j(y)}, \nonumber \\
   &&\left[ K_{a}^{i}(x), K_{b}^{j}(y)  \right]_D^{} =\left[\kappa_{a}^{i}(x),\kappa_{b}^{j}(y)\right]_D^{}
  =\left[\kappa_{a}^{i}(x),\kappa_{b}^{j}(y)\right]~\nonumber \\
  &&~~~~~~~~~~~~~~~~~~~~~~~~~~~~~~~~~~~~~~~~~~~~~=\left[A_{a}^{i}(x),f_{b}^{i}(y)\right]~+~
  \left[f_{a}^{i}(x),A_{b}^{j}(y)\right ]=0~~~~~~~\label{Dirac1'}
\eeq
where in the last terms of second and third equations, the Poisson brackets are to be evaluated using 
(\ref{A3}) which express $f^i_a(E)$ as functions of $E_a^i$.

 The Dirac brackets of $(A^{i}_{a},
\hat{E}^{a}_{i})$ and $(\hat{E}^{a}_{i},\hat{E}^{b}_{j})$ are not same
as their Poisson brackets:
\begin{eqnarray}
 \left[A_{a}^{i}(x),~\hat{E}^{b}_{j}(y)\right]_D^{}  &= &\left[A_{a}^{i}(x),~E^{b}_{j}(y)-\frac{2}{\eta}~\tilde{e}^{b(\eta)}_{0j}(y)\right]_D^{} 
   =\delta^{b}_{a} \delta^{i}_{j} ~\delta^{3}(x,y)-
  \frac{2}{\eta}~\left[A_{a}^{i}(x), ~\tilde{e}^{b(\eta)}_{0j}(\kappa;y)\right],   \nonumber \\
 \left[{\hat E}^a_i(x),~\hat{E}^{b}_{j}(y)\right]_D^{} &=& 
   \frac{4}{\eta^2}~
   \left[\tilde{e}^{a(\eta)}_{0i}(\kappa;x),~\tilde{e}^{b(\eta)}_{0j}(\kappa;y)\right]  \nonumber \\
   &=& -~\frac{4(\th^2_{}+\phi^2_{})}{\eta^3_{}}~ \e^{acd}_{} \e^{bef}_{} 
   \left(\left[F^i_{cd}(x), ~\e^{jmn}_{} \kappa^m_e(y) \kappa^n_f(y) \right] \right. \nonumber \\
   &&~~~~~~~~~~~~~~~~~~~~~~~~~~~~~~~~~~~~~~~\left.+ \left[\e^{imn}_{} \kappa^m_c(x) \kappa^n_d(x),~ F^j_{ef}(y) \right] \right) \nonumber \\
   &=& \frac{4(\th^2_{}+\phi^2_{})}{\eta^2_{}}~ \e^{acd}_{} \e^{bef}_{} 
   \left[ D^{}_{[c}(A)\kappa^i_{d]}(x), ~D^{}_{[e}(A)\kappa^j_{f]}(y) \right] 
    \label{Dirac2}
\end{eqnarray}
Here the argument $\kappa$ in $e^a_{0i}(\kappa)$ and ${\tilde e}^a_{0i}(\kappa)$  is to indicate that 
these are as in (\ref{e(A,K)}) with $K^i_a$ replaced by $\kappa^i_a$ which in turn are given by (\ref{A3})
as functions of $A^i_a$ and $E^a_i$. Further, here in the second equation, we have used:
\beq
\left[E^{a}_{i}(x), ~{\tilde e}^{b(\eta)}_{0j}(\kappa;y)\right]
  ~ + ~\left[\tilde{e}^{a(\eta)}_{0i}(\kappa;x), ~E^{b}_{j}(y)\right]~=~0 \nonumber
\eeq
Also,
\beq
&&\left[A_{a}^{i}(x),~\hat{F}^{b}_{j}(y)\right]_D^{}=\frac{2(1+\eta^2)}{\eta}\left[A_{a}^{i}(x),~\tilde{e}^{b}_{0j} (y)\right]_D 
 =\frac{2(1+\eta^2)}{\eta}~\left[A_{a}^{i}(x),~\tilde{e}^{b}_{0j}(\kappa;y)\right]  \nonumber\\
&&\left[E^{a}_{i}(x),~\hat{F}^{b}_{j}(y)\right]_D^{}=
\frac{2(1+\eta^2)}{\eta}[E^{a}_{i}(x),~\tilde{e}^{b}_{0j}(y)]_D=\frac{2(1+\eta^2)}{\eta}\left[E^{a}_{i}(x),~\tilde{e}^{b}_{0j}(\kappa;y)\right] \nonumber \\
&& \left[{\hat F}^a_i( x), ~{\hat F}^b_j( y)\right]_D=\frac{4(1+\eta^2_{})^2_{}}{\eta^2_{}}\left[ {\tilde e}^a_{0i}( x),~ {\tilde e}^b_{0j}(y) \right]_D = 
\frac{4(1+\eta^2_{})^2_{}}{\eta^2_{}}\left[ {\tilde e}^a_{0i}(\kappa;x),~ {\tilde e}^b_{0j}(\kappa;y) \right]\nonumber \\
&&~~~~~~~~~~~~~~~~~~~~~~= \frac{4(1+\eta^2_{}) }{\eta^2_{}}
\left( \th^2_{} +\phi^2_{}\right) \e^{acd}_{} \e^{bef}_{} 
   \left[ D^{}_{[c}(A)\kappa^i_{d]}(x), ~D^{}_{[e}(A)\kappa^j_{f]}(y) \right] \nonumber \\
&& \left[{\hat E}^a_i(x), {\hat F}^b_j(y)\right]_D + \left[{\hat F}^a_i(x), {\hat E}^b_j(y)\right]_D 
  = -~\frac{8\left(1+\eta^2_{}\right)}{\eta^2_{}} \left[{\tilde e}^a_{0i}(\kappa; x),~ {\tilde e}^b_{0j}(\kappa; y)\right]  \nonumber \\ 
  &&~~~~~~~~~~~~~~~~~~~~~=  -~\frac{8\left( \th^2_{} +\phi^2_{}\right)}{\eta^2_{}}   \e^{acd}_{} \e^{bef}_{} 
   \left[ D^{}_{[c}(A)\kappa^i_{d]}(x), ~D^{}_{[e}(A)\kappa^j_{f]}(y) \right]  \nonumber  \\
   && \left[ K^i_a(x), ~{\hat F}^b_j(y)\right]_D = \frac{2(1+\eta^2_{})}{\eta} \left[K^i_a(x),
   ~{\tilde e}^b_{0j}( y)\right]_D= \frac{2(1+\eta^2_{})}{\eta} \left[\kappa^i_a(x),
   ~{\tilde e}^b_{0j}(\kappa;y)\right]  
~~~~~~~~~ ~\label{Dirac3}
\eeq

\end{document}